\documentclass[onecolumn,amsmath,amssymb,nofootinbib,12pt]{article}
\usepackage{jheppub}
\usepackage{ifpdf}
\usepackage{graphicx,subcaption}
\usepackage{amsfonts}
\usepackage{amsmath}
\usepackage{amssymb}
\usepackage{epsfig}
\usepackage{pdfpages}
\usepackage{graphicx,epstopdf}
\usepackage[makeroom]{cancel}
\hypersetup{pdftex,colorlinks=true,allcolors=blue}
\usepackage{array}
\usepackage{ulem}
\usepackage{bm}

\newcommand{\fig}[1]{Fig.\ref{#1}}
\newcommand{\bra}[1]{|#1\rangle}
\def\be{\begin{equation}}
\def\ee{\end{equation}}
\def\ba{\begin{eqnarray}}
\def\ea{\end{eqnarray}}

\def\nn{\nonumber}
\def\lf{\left}
\def\rt{\right}

\newcommand{\eq}[1]{(\ref{#1})}

\def\nn{\nonumber}\def\lf{\left}\def\rt{\right}\def\q{\theta} \def\w{\omega}  \def\y {\psi}   \def\p {\pi} \def\a {\alpha} \def\s {\sigma} \def\d {\delta} \def\f {\phi} \def\g {\gamma}   \def\k {\kappa} \def\l {\lambda}  \def\x {\xi}  \def\b {\beta}  \def\m {\mu} \def\pd {\partial} \def \inf {\infty}  
 \def\W{\Omega} \def\Y {\Psi}     \def\D {\Delta}   \def\L {\Lambda} \def\X {\Xi}   \def\.{\cdot}
\def\math {\mathcal}
\usepackage{xspace}
\begin{document}
\title{\Large Circuit Complexity for Fermionic Thermofield Double
States}
\author[a]{Jie Jiang,}
\author[a]{Xiangjing Liu}
\affiliation[a]{Department of Physics, Beijing Normal University,
Beijing 100875, China}
\emailAdd{jiejiang@mail.bnu.edu.cn}
\emailAdd{liuxj@mail.bnu.edu.cn}
\date{\today}
\abstract{Motivated by the holographic complexity proposals, in this paper, we investigate the time dependence of the complexity for the Fermionic thermofield double state (TFD) using the Nielsen approach and Fubini-Study (FS) approach separately. In both two approaches, we discuss the results for different reference states:  the Dirac vacuum state and the Gaussian state which has no spatial entanglement (NSE). For Dirac vacuum reference state, we find that the complexity by both two approaches is time independent and the circuit complexity shares the same expression for both methods with the $L^2$ norm. For the NSE reference state, the complexity by the Nielsen approach is time-dependent while it by the FS approach is time-independent.
 Further, we find our dynamical results are in good agreement with the bosonic case, where the complexity evolves in time and saturates after a time at the order of the inverse temperature. And we show that the complexity of formation is also shared same similar behaviors with the holographic complexity.
}
 \maketitle
\section{Introduction}
Recently, applications of quantum information concepts to high energy physics and gravity have led to many promising results. Among all, it has become more clear that special properties of entanglement in holographic quantum field theories are account for the emergence of smooth higher-dimensional geometries in gauge-gravity duality\cite{Maldacena:2013xja,Aharony:1999ti,Takayanagi:2016,VanRaamsdonk:2010pw}. This faith inspired a further deeper idea that ``ER=EPR''\cite{Maldacena:2013xja} when consider a  non-traversable wormhole created by an Einstein-Rosen (ER)\cite{Einstein:1935} bridge and a pair of maximally entangled black holes, based on the point of view that a black hole might be highly entangled with a system that is effectively infinitely far away. EPR denotes the quantum entanglement (Einstein-Podolsky-Rosen paradox). However, recent developments pointed out that holographic entanglement is unable to describe the bulk spacetime far behind the event horizon of black holes \cite{Susskind:2014moa}. To get more insights into the procedure of sending a signal through ER bridge, a new quantity named ``complexity'', originated from the field of quantum computations, was introduced into the context of holographic gravity. In essence, the complexity presents the minimum number of elementary operations or gates from a reference state to a target state.

From the perspective of quantum entanglement in holography, thermofield double (TFD) states play a crucial role. TFD states can be constructed by entangling two copies of conformal field theory (CFT),  in this way that integrating out either copy induces the thermal density matrix at inverse temperature $\beta>0$ for the other \cite{BrD,Maldacena:2003}, $i.e.$,
\ba\label{TFD}\begin{aligned}
\bra{TFD(t_L,t_R)}=Z_{\beta}^{-1/2}\sum_n e^{-\beta E_n/2} e^{-iE_{n}(t_L+t_R)}\bra{E_n}_L\bra{E_n}_R\,,
\end{aligned}
\ea
where $\bra{E_n}_{L,R}$ denotes the energy eigenstate of the left/right CFT, and $Z_{\beta}$ denotes the partition function at inverse temperature $\beta$. In the context of holographic gravity,  the TFD states  \eq{TFD} dual to the left and right sides of the geometry  which is connected by a wormhole, or ER bridge, and the volume of this geometry increases for a time which is exponential in the number of degrees of freedom of the boundary theory. However, this time is much larger than other characteristic times in holographic gravity which indicates that entanglement alone is not enough to capture the dynamics behind the horizon and there must exist some other quantity we yet do not know which have the information about this late-time evolution of the wormhole interior.

\begin{figure}[t]
\centering
\includegraphics[width=0.45\textwidth]{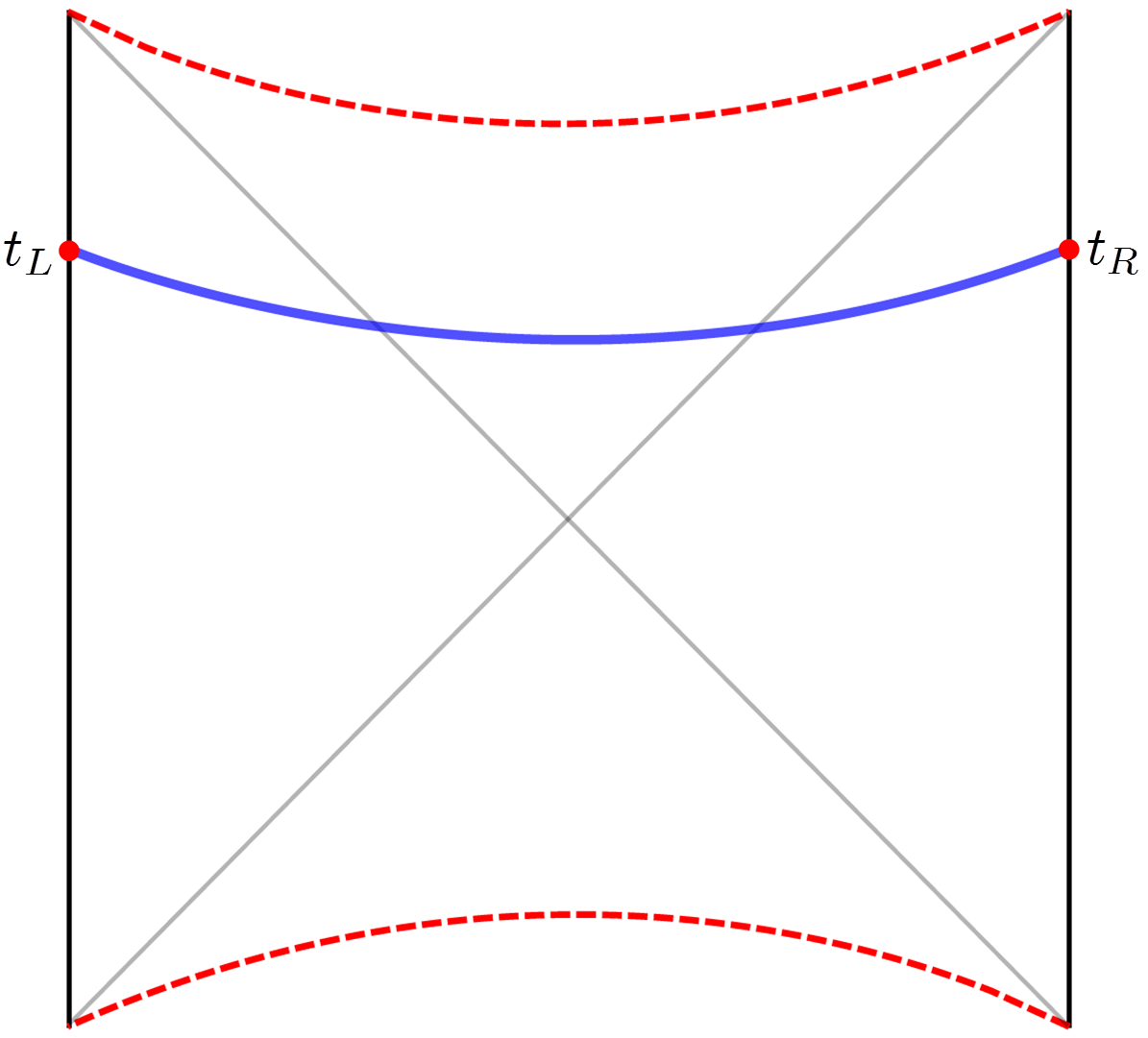}
\ \ \ \ \
\includegraphics[width=0.45\textwidth]{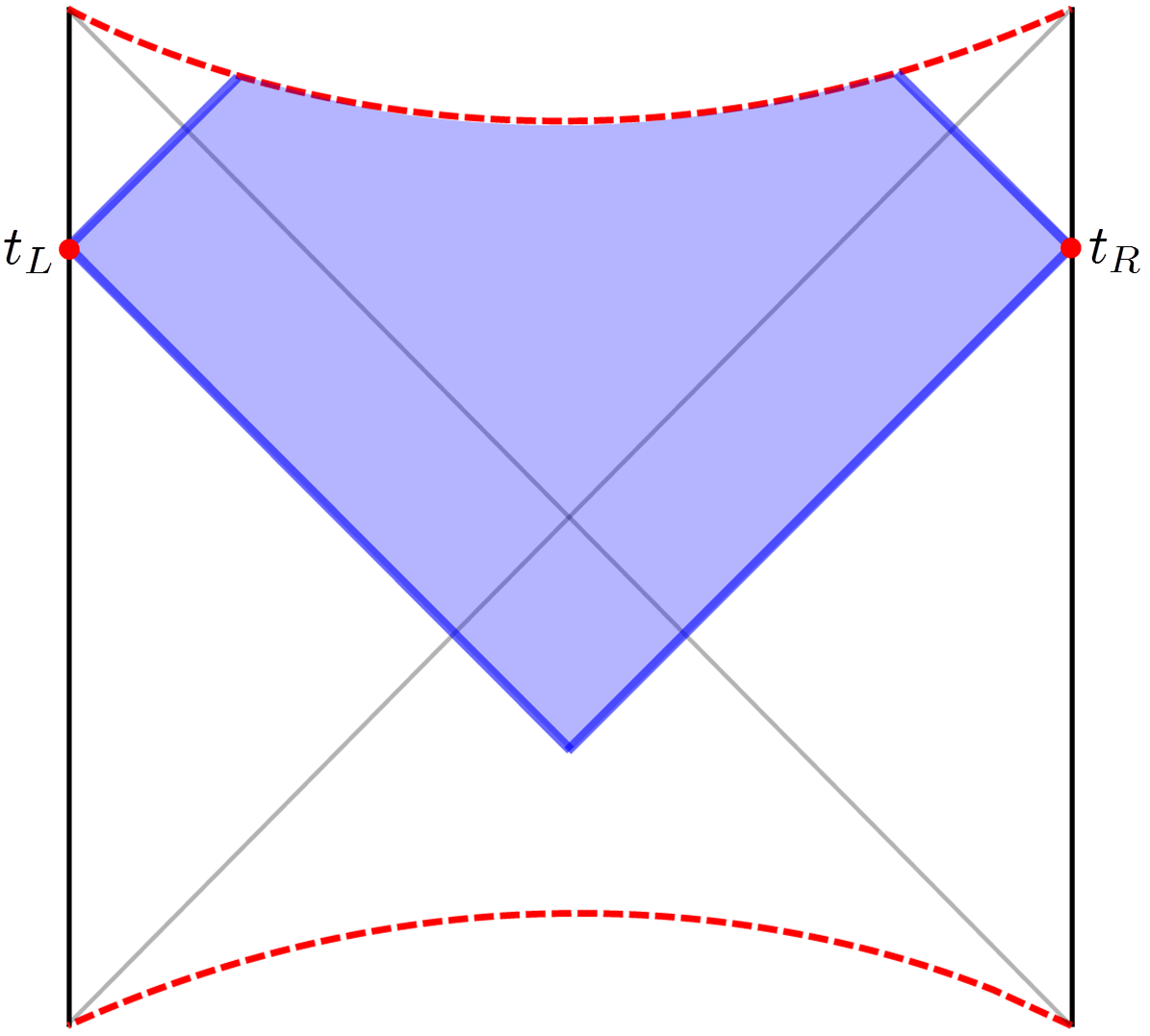}

\caption{Complexity=volume (CV, left) and complexity=action (CA, right) for the eternal AdS black hole dual to the TFD state.  The blue curve in the left panel denotes a spacelike surface which has maximal volume that connects the specified time slices on the left and right boundaries. The shaded region in the right panel is the corresponding Wheeler-DeWitt patch.}\label{CVCA}
\end{figure}

It has been suggested by Susskind and collaborators that this quantity is the quantum computational complexity of the boundary state. Particularly, they came up with two proposals for quantifying the size of ER bridge dual to holographic complexity: complexity=volume(CV) \cite{1,2} and complexity=action(CA) \cite{BrL,BrD}. The CV conjecture suggests that the complexity of the boundary state is dual to the volume of an extremal codimension-one surface extending the boundary time slice into the bulk. The CA conjecture believes that the complexity of the boundary state is proportional to the gravitational action of the bulk region known as the Wheeler-deWitt (WdW) patch bounded by light sheets. One can get a more visual image about the CV and CA conjectures in the \fig{CVCA}.  And those two conjectures have been vigorously investigated in the most recent literatures\cite{A1,A2,A3,A4,A5,A6,A7,A8,A9,A10,A11,A12,A13,A14,A15,A16,A17,A18,A19,A20,A21,A22,A23,A24,A91,A51}.

 However, those research program is still in a very primary stage due to the fact that we still lack a precise definition of complexity in the context of the boundary CFT, or more generally in quantum field theories. However, some initial steps towards developing such a precise definition have been taken in the past few years \cite{Brown:2018BS,Caputa:2017,Yang:2018nda,Hashimoto:2017,Jefferson:2017,Chapman:2017,Yang:2017,Yang:2018,Yang2018YA,Reynolds:2017,Khan:2018,Guo:2018kzl,Jiang:2018gft,
Chapman:2018hou,Hackl:2018ptj}. In particular, ref. \cite{Hackl:2018ptj} adapted the approach of Nielsen and his collaborators \cite{Nielsen1,Nielsen2,Nielsen3} to translate the task of finding the complexity of the ground state of a free fermionic quantum field theory into a geometric problem of finding optimal geodesics in an associated geometry. And also, based on the Fubini Study metric, there is a similar geometric approach proposed in \cite{Chapman:2017,Yang:2018} to deal with this question. Previous results \cite{Hackl:2018ptj,Chapman:2017,Jefferson:2017} have revealed some very interesting similarities with holography in the structure of the UV divergences. Although the correctness of the CV or CA proposals are still vague, it may provide us with more insights.

In this paper, we are extending the construction of ref. \cite{Hackl:2018ptj} to evaluate the complexity of the TFD state in a free fermionic quantum field theory. That is, the state \eq{TFD} will serve as our target state, while we will start from different reference states. It is noticeable that the quadratic operators of the canonical operators are sufficient to generate circuits in interest. The key feature which permits the construction of circuits is that both the reference and target states are Gaussian states. Furthermore, all of the intermediate states will be Gaussian as well due to all the gates are generated by the quadratic operators. In the following, we will use two approaches to manipulate these Gaussian states, one is based on the covariance matrix, the other is based on the Fubini-Study metric. Moreover, by applying the Fubini-Study approach, we will investigate circuits from two different reference states. These analyses will allow us to study the time-dependence of complexity, as well as the complexity of formation of the thermal state, and to compare our results with those of the holographic complexity conjectures. However, for fermionic TFD state, we show that the complexities, as well as the optimal circuits, derived by Nielsen and Fubini-Study approaches are different.

 This work is organized as follows: In section \ref{Pre}, we introduce the required preliminaries. We review some key elements of Nielsen's approach to evaluate circuit complexity, and then demonstrate how both the time-independent and time-dependent fermionic TFD states of $2N$ decoupled harmonic oscillators can be generated by quadratic operators, also for later use, we write down the covariance matrices, which serves as an alternate--and indeed, more natural--characterization of Gaussian states in the last part of this section. In section \ref{CofTs}, We then proceed to evaluate the complexity for TFD states comprised of $N$ pairs of harmonic oscillators as well as TFD states in free Dirac field theory. In section \ref{FSmetric}, we adopt the Fubini-Study metric approach to investigate the circuit complexity for TFD states, moreover we also study the spatial unentangled reference state. In section \ref{conclusion}, we conclude this work.
 
\section{Preliminaries: fermionic thermofield double states and covariance matrix}\label{Pre}
In this section, we provide the relevant preliminaries for the construction of both the fermionic TFD states and the relevant quantum circuits. We start with the construction of the fermionic thermofield double states for 2$N$ simple harmonic oscillator in preparation for our studies of its complexity in the rest of the paper. Then for later convenience we write down the covariance matrix form of the Gaussian states.

\subsection{TFD state for the simple fermionic harmonic oscillator}
  As demonstrated in the introduction, the purpose of present work is to expand the investigation of the complexity of the free bosonic TFD state to the fermionic case. we will follow the same approach introduced in ref.\cite{Hackl:2018ptj} to consider the situation. However we should build the free fermionic TFD state first. In this subsection, we explicitly demonstrate the steps to construct the TFD state for the free fermionic QFT from two identical copies of a simple harmonic oscillator.

Let us introduce some basic notation before we proceed. We denote by subscripts $L$ and $R$ in analogy to the left and right copies of the CFT in the Penrose diagram in \fig{CVCA}. The whole Hilbert space of the system can be constructed by the tensor product of those two separated copies $\math{H}=\math{H}_L\otimes \math{H}_R$. There are two pairs of creation and annihilation operators $\{(a_L,a_L^\dag),(a_R,a_R^\dag)\}$  for the left and right subsystem which satisfy the non-vanish anti-commutation relations $\{a_L,a_L^\dag\}=\{a_R,a_R^\dag\}=1 $. Then, in terms of the canonical creation and annihilation operators, the Hamiltonian of this system is written by
\ba
H=H_L+H_R=\w \lf(a_L^\dag a_L+a_R^\dag a_R\rt)\,,
\ea
where $\omega$ denotes the frequency of the oscillator and since the constant term is irrelevant we have already dropped it.

The normalized energy eigenstates are given by acting on the vacuum $\bra{0}$ with creation operators. Considering that we are discussing the fermionic QFT, the total Hilbert space is given by
\ba
\math{H}=\text{span}\{\bra{0}_L\otimes\bra{0}_R,\bra{1}_L\otimes\bra{0}_R,\bra{0}_L\otimes\bra{1}_R,\bra{1}_L\otimes\bra{1}_R\},
\ea
  where $\bra{1}_{\L}=a_{\L}^\dag\bra{0}_{\L}$ for $\L=\{L,R\}$. The excited state $\bra{1}_{\Lambda}$ is the eigenstate of system $H_{\L}$ which means
\ba
H_{\L}\bra{1}_{\L}=\w \bra{1}_{\L}\,.
\ea
Also note that one important observation is the TFD state \eq{TFD} is invariant when we shift
\ba
t_L\to t_L+\Delta t,\ \ \ \ \ t_R\to t_R-\Delta t\,.
\ea
$i.e.$, the TFD state is invariant if the time evolve with the combined Hamiltonian $H_L-H_R$. As a result, the holographic complexity only depends on the combination $t=t_L+t_R$ and it remains unchanged by above shifts. In fact, this invariance is encoded in the `boost symmetry' of the dual black hole geometry. Hence, in the following we begin with the fermionic TFD state at the time $t=0$, then we consider the generalization to the time dependent TFD state.

\subsubsection{TFD state at $t=0$}

  In this subsection, we consider the case of the  fermionic TFD state at $t_L+t_R=0$ which can be constructed from two copies of the vacuum state by acting with the creation operators  in the following manner:
\ba\begin{aligned}\label{TFDt0}
\bra{TFD}&=\lf(1+e^{-\b \w}\rt)^{-1/2}\sum_{n=0}^{1}e^{-n\b \w/2}\bra{n}_L \bra{n}_R\\
&=\lf(1+e^{-\b \w}\rt)^{-1/2}\exp\lf[e^{-\b\w/2}a_L^{\dag}a_R^{\dag}\rt] \bra{0}_L \bra{0}_R,
\end{aligned}\ea
where we have taken the $E_n=n\omega$ and omitted the tensor product symbol for simplicity. As previously mentioned, we  adopt the Neilsen's geometric approach to evaluate the circuit complexity and we wish the circuit is constructed from unitary operators. Thus, we next try to rewrite the fermionic TFD state \eq{TFDt0} by acting with a unitary operator on the vacuum $\bra{0}_L\bra{0}_R$. Indeed, the express \eq{TFDt0} can be written in terms of an unitary transformation through following steps.

  First, we define
\ba\begin{aligned}
u(\beta)&=\cos\q_0=\frac{1}{\sqrt{1+e^{-\b\w}}},\,  \text{and}\\
v(\beta)&=\sin\q_0=\frac{e^{-\b\w/2}}{\sqrt{1+e^{-\b\w}}},
\end{aligned}\ea
such that
\ba
u^2(\beta)+v^2(\beta)=\cos^2\theta_0+\sin^2\theta_0=1.
\ea

  Then, according to the anti-commutation relations of creation and annihilation operators, the fermionic TFD state at $t=0$ can be written as
\ba
\bra{TFD}=\lf(\cos\q_0+\sin\q_0 a_L^{\dag}a_R^{\dag}\rt)\bra{0}_L\bra{0}_R\, .
\ea

   Finally, using the property
\ba\begin{aligned}
(a_L^{\dag}a_R^{\dag}+a_La_R)^{2n}\bra{0}_L\bra{0}_R&=(-1)^n\bra{0}_L\bra{0}_R,\\
(a_L^{\dag}a_R^{\dag}+a_La_R)^{2n+1}\bra{0}_L\bra{0}_R&=(-1)^n\bra{1}_L\bra{1}_R,
\end{aligned}\ea
we obtain the desired result
\ba
\bra{TFD}=U(\beta)\bra{0}_L\bra{0}_R\,,
\ea
where $U(\beta)=e^{\q_0 \hat{K}_0}$ is a unitary operator with
\ba\label{q0}
\q_0=\arctan (e^{-\b\w/2})\,
\ea
and
\ba
\hat{K}_0=a_L^{\dag}a_R^{\dag}+a_La_R.
\ea
Note that the ordering of the operators in these expressions is crucial since the anti-commutation relations among them.

At this point, the set-up for the time-independent ($t=0$) fermionic TFD is complete. In favor of evaluating the circuit complexity using the covariance matrix approach suggested in ref. \cite{Chapman:2018hou}, we define a set of Hermitian fermionic operators given by
 \ba\begin{aligned}\label{cmQP}
Q_{\L}&=\frac{1}{\sqrt{2}}\lf(a_{\L}^{\dag}+a_{\L}\rt)\,\\
P_{\L}&=\frac{i}{\sqrt{2}}\lf(a_{\L}^{\dag}-a_{\L}\rt)\,,
\end{aligned}\ea
which are commonly called Majorana modes. In contrast to the analogous bosonic operators, they do not consist of conjugate pairs $(Q_{\Lambda},P_{\Lambda})$, but rather they are governed by the anti-commutation relations, and one can straightforward to verify that these operators satisfy
\ba\begin{aligned}
\{Q_{\L}, Q_{\L'}\}&=\{P_{\L}, P_{\L'}\}=\d_{\L\L'}\,,\\
\{Q_{\L}, P_{\L'}\}&=0\,.
\end{aligned}\ea
Then, in terms of the Majorana modes, we re-express the generator as
\ba
\hat{K}_0=Q_L Q_R-P_L P_R\,.
\ea
Now, it is more clear that the fermionic TFD state is entangled after the action of the unitary operator $U(\beta)=e^{\theta_0(Q_L Q_R-P_L P_R)}$ on the unentangled ground $\bra{0}_L\bra{0}_R$. Hence, in this case, the unitary operator $U(\beta)$ can be interpreted as an operator which entangles the left and right subsystem. This also further demonstrates that the fermionic TFD state at time $t=0$ is entangled. We can obtain the thermal density matrix for the other after tracing out one copy in this way.

\subsubsection{Time-dependent TFD state}
In this subsection, we follow the steps above to extend the construction to the time-dependent fermionic TFD state. For simplicity, we shall follow the common convention in holography (see, $e.g.$, refs. \cite{Stanford:2014jda,Carmi:2017jqz}) and set $t_L=t_R=t/2$. Then, the time-dependent fermionic TFD state is given by
\ba\begin{aligned}
\bra{TFD(t)}&=\lf(1+e^{-\b \w}\rt)^{-1/2}\sum_{n=0}^{1}e^{-n\b \w/2}e^{-in\w t}\bra{n}_L\bra{n}_R\\
&=\lf(1+e^{-\b \w}\rt)^{-1/2}\exp\lf[e^{-\b\w/2-i\w t}a_L^{\dag}a_R^{\dag}\rt]
\bra{0}_L\bra{0}_R\\
&= \lf(\cos\q_0+\sin\q_0 e^{-i \w t} a_L^{\dag}a_R^{\dag}\rt)\bra{0}_L\bra{0}_R
\end{aligned}\ea
where we have used $E_n=n\omega$ and $\theta_0$ is defined in \eq{q0} . As illustrated above, we also wish to express this time-dependent state in terms of a unitary operator acting on the vaccum state.

 Similarly, we observe that
\ba\begin{aligned}
\lf(e^{-i\w t} a_L^{\dag}a_R^{\dag}+e^{i\w t} a_La_R\rt)^{2n}\bra{0}_L\bra{0}_R&=(-1)^n\bra{0}_L\bra{0}_R\,,\\
\lf(e^{-i\w t}a_L^{\dag}a_R^{\dag}+e^{i\w t}a_La_R\rt)^{2n+1}\bra{0}_L\bra{0}_R&=(-1)^n e^{-i\w t}a_L^{\dag}a_R^{\dag}\bra{0}_L\bra{0}_R\,.
\end{aligned}\ea
Then, by applying this observation, one can obtain the desired result
\ba\label{TFDt}
\bra{TFD(t)}=e^{z a_L^{\dag}a_R^{\dag}+z^*a_La_R}\bra{0}_L\bra{0}_R=U(\beta,t)\bra{0}_L\bra{0}_R\,,
\ea
where $U(\beta,t)=e^{z a_L^{\dag}a_R^{\dag}+z^*a_La_R}=e^{\theta_0\hat{K}(t)}$ is a unitary operator with
\ba\begin{aligned}
z&=\q_0 e^{-i\w t},\\
\hat{K}(t)&=e^{-i\w t}a_L^{\dag}a_R^{\dag}+e^{i\w t}a_La_R.
\end{aligned}\ea
 Similarly, for later convenience, we can re-express the generator in terms of the Majorana modes as
\ba\label{mmkt}
\hat{K}(t)=\lf(Q_L Q_R-P_L P_R\rt)\cos\w t-\lf(P_L Q_R+Q_L P_R\rt)\sin\w t\,.
\ea
Then, we have
\ba
\bra{TFD(t)}=e^{\q_0\hat{K}(t)}\bra{0}_L\bra{0}_R\,,
\ea

It implies that the time involution of the TFD state can be interpreted as the entanglement of the Majorana modes $P$ and $Q$.
\subsection{Covariance matrix}
Most recently, researchers have made two equivalent descriptions of the Gaussian states of the bosonic system \cite{Chapman:2018hou} to investigate the circuit complexity for TFD states in free scalar quantum field theories, one is in terms of their wavefunction, and the other is in terms of the covariance matrix. They found that the covariance matrix representation is much simpler than the former. To learn from them and also avoid the unnecessary works, we will skip the wavefunction representation straightforward to the covariance matrix approach to study the circuit complexity of the fermionic TFD states for  harmonic oscillators in the present work. In this subsection, we will review the relevant aspects of this approach based on ref.\cite{Hackl:2018ptj}.

A Fermionic system with $N$ degrees of freedom can be described by $2N$ linear Majorana modes $\x^a\equiv(Q_1,\cdots,Q_N,P_1,\cdots,P_N)$. As demonstrated above, these pairs $(Q_i,P_i)$ are governed by the anti-commutation relations: $\{Q_i,Q_j\}=\delta_{ij}=\{P_i,P_j\}$ and $\{Q_i,P_j\}=0$. For a general Gaussian state $\bra{\psi}$, their covariance matrix can be expressed by
\ba\label{Wab}
\langle\Y|\x^a\x^b\bra{\Y}=\langle\Y|\x^{[a}\x^{b]}\bra{\Y}+\langle\Y|\x^{(a}\x^{b)}\bra{\Y}=\frac{1}{2}\lf(G^{ab}+i\W^{ab}\rt)\,.
\ea
where $G^{ab}$ is the symmetric part of the correlation matrix and $\Omega^{ab}$ denotes the antisymmetric part.
Consider a pure fermionic Guassian state ,$i.e.$, $\langle\psi|\x^a\bra{\psi}=0$, the symmetric component of the covariance matrix is fixed by
\ba\label{antc}
G^{ab}=\{\x^a,\x^b\}=\d^{ab}\, ,
\ea
while the antisymmetric part $\Omega^{ab}$ completely characterizes the fermionic Gaussian state $\bra{\psi}$. Hence we can use $\Omega$ as label for these Gaussian states without ambiguity.

   Now, we consider the fermionic TFD state with the degree of the freedom $N$ for the left and right subsystem with the Majorana mode
   \ba
    \x^a\equiv\lf(Q_{L1},Q_{R1},P_{L1},P_{R1},\cdots,Q_{LN},Q_{RN},P_{LN},P_{RN}\rt).
    \ea
  The Hilbert space of this decoupled system with total degrees of freedom $2N$ is given by
   \ba
    \math{H}=\otimes_I \math{H}^I=\otimes_{I}\lf(\math{H}_L^{I}\otimes \math{H}_R^{I}\rt) \ \ \text{with} \ \ I=(1,\cdots,N),
    \ea
    where $\math{H}_{\L}^{I}$ denotes the Hilbert space of a Fermionic harmonic oscillator labeled by $I$ in the subsystem $\Lambda$. In particular,  the Gaussian state corresponding to Majorana mode $\x^a$ in this Hilbert can be expressed as
\ba
\bra{\W_0}=\otimes_I \bra{\W_0^{(I)}}=\otimes_I \lf(\bra{0}^I_{L}\otimes\bra{0}^I_{R}\rt)\,.
\ea
Using \eq{Wab}, it is straightforward to obtain
\ba\label{W0}
\W_0=\oplus_I \W^{(I)}_0\,.
\ea
with
\ba
\W^{(I)}_0=
\left(
\begin{array}{cccccccc}
 0 & 0 & \ \, 1\ \, & 0 \\
 0 & 0 & 0 &  \ \, 1\ \, \\
 -1 & 0 & 0 & 0 \\
 0 & -1 & 0 & 0 \\
\end{array}
\right)\,.
\ea

 Next, we wish to construct a circuit  from $\bra{\Omega_0}$ regarded as the reference to the target state $\bra{\Omega}$ by a unitary transformation $U$ corresponding to a specific Bogoliubov transformation, $i.e.$,
\ba
\bra{\W}=\hat{U}\bra{\W_0}\,.
\ea

Hence, we consider the corresponding action on the operator $\x^a$, $i.e.$,
\ba\label{Uab}
\hat{U}^{\dag}\x^a\hat{U}=U^a{}_b\x^b\,.
\ea
As relevant analysis given in ref.\cite{Hackl:2018ptj} reveals, the desired transformation U is an element of the rotation group $SO(4N)$.
Then, the covariance matrix of the target state $\bra{\W}$ can be computed as
\ba\begin{aligned}\label{covm}
\W^{ab}&=-i\langle\W_0|U^{\dag}\lf[\x^{a},\x^{b}\rt]U\bra{\W_0}=U^a{}_c\W_0^{cd}U^b{}_d=\lf(U\W_0U^T\rt)^{ab}\,,
\end{aligned}\ea
which implies that the key to compute the covariance matrix $\W$ is to obtain the matrix $U$.

 Thus, to evaluate the circuit complexity from the vacuum state $\bra{\Omega_0}$ to the time-dependent fermionic TFD state $\bra{\text{TFD}(t)}$,  we need the full information about this unitary transformation $U$ to obtain the covariance matrix of $\bra{\text{TFD}(t)}$.  Fortunately, In the above subsection we managed to obtain a exponential form $U=e^{\theta_0 \hat{K}(t)}$. The matrix form of this is given in the appendix A. The next step is using the matrix $U$ to obtain the covariance matrix of $\bra{\text{TFD}(t)}$ with $2N$ degrees of freedom.

  Consider the time-dependent fermionic TFD state $\bra{\text{TFD}(t)}$ which in this case can be expressed as
\ba\label{trajectory}
\bra{TFD(t)}=\otimes_I \bra{TFD(t)}_I=\otimes_I \left[U_I\bra{\Omega_0}_I\right]=\otimes_I\lf[ e^{\q_0^{(I)}\hat{K}_I(t)}\bra{\W_0}_I\rt]\,,
\ea
in which
\ba\begin{aligned}
\q_0^{(I)}&=\arctan e^{-\b\w_I/2}\,,\\
\hat{K}_I(t)&=\lf(Q_L Q_R-P_L P_R\rt)\cos\w_I t-\lf(P_L Q_R+Q_L P_R\rt)\sin\w_I t\,.\\
\end{aligned}\ea
Then, by the virtue of Eqs.\eq{Uq} and \eq{covm}, the covariant matrix of this TFD state can be expressed as
\ba\label{WTq}
\W_T(t)=\oplus_I\W_T^{(I)}\,.
\ea
with
\ba
\W_T^{(I)}=
\left(
\begin{array}{cccc}
 0 & \sin 2 \theta_0^{(I)}  \sin \omega_I t & \cos2 \theta_0^{(I)}  & \cos \omega_I t \sin 2 \theta_0^{(I)}  \\
 -\sin 2 \theta_0^{(I)}  \sin  \omega_I t & 0 & -\cos \omega_I t \sin 2 \theta_0^{(I)}  & \cos 2 \theta_0^{(I)}  \\
 -\cos 2 \theta_0^{(I)} & \cos  \omega_I t \sin 2 \theta_0^{(I)}  & 0 & -\sin 2 \theta_0^{(I)}  \sin  \omega_I t \\
 -\cos \omega_I t \sin 2 \theta_0^{(I)}  & -\cos 2 \theta_0^{(I)}  & \sin 2 \theta_0^{(I)}  \sin  \omega_I t & 0 \\
\end{array}
\right)\,.
\ea

 Finally, suggested by covariance matrix approach, the invariant information about the relation between the reference state $\bra{\Omega_0}$ and the target state $\bra{\Omega}$ is encoded in the relative covariance matrix
  \ba
  \Delta^a{}_b=\Omega^{ac}\omega^{(0)}_{cb} \ \ \ \ \text{with} \ \ \omega^{(0)}=\Omega_0^{-1},
  \ea
$i.e.$, $\Omega_{0}^{ac}\omega^{(0)}_{cb}=\delta^a{}_b$. More specifically, the information of circuit complexity is captured in the eigenvalues of this matrix.

\section{Complexity of fermionic TFD states }\label{CofTs}

\subsection{Circuit complexity of harmonic oscillators}
  In this section, we consider the circuit complexity from the unentangled vacuum state $\bra{\psi_R}=\bra{\Omega_0}$ to the target state $\bra{\psi_T}=\bra{\text{TFD}(t)}$ for free fermionic harmonic ascillators using the covariance matrix approach. As mentioned in the above section, in Nielsen's approach to evaluate the circuit complexity, one should choose a particular cost function equivalent to given a bi-invariant metric. Then, the circuit complexity for a specific cost function directly connects to the relative covariant matrix, $i.e.$ $\math{C}(\bra{\W_R}\to\bra{\W_T})=||\hat{K}||=||\log\Omega_T \omega_R||/2$. In this case, It can be expressed by
 \ba
\math{C}(\bra{\W_0}\to\bra{\text{TFD}(t)})=||\theta_0^{(I)}\hat{K}_I(t) ||=\frac{||\log\Omega \omega^{(0)}||}{2}.
\ea

 Then,  to calculate this circuit complexity, one should obtain the relative covariance matrix. According to \eq{W0} and \eq{WTq}, the relative covariant matrix can be expressed as
\ba
\D=\oplus_I\D_I\, \ \ \text{with} \ \ \D_I=(\Omega)_I (\omega_0)_I
\ea
with
\ba
\D_I=\left(
\begin{array}{cccc}
 \cos2 \theta_0^{(I)} & \cos \omega_I t \sin 2 \theta_0^{(I)} & 0 & -\sin  \omega_I t \sin 2 \theta_0^{(I)} \\
 -\cos \omega_I t \sin 2 \theta_0^{(I)} & \cos 2 \theta_0^{(I)} & \sin \omega_I t \sin 2 \theta_0^{(I)} & 0 \\
 0 & -\sin \omega_I t \sin2 \theta_0^{(I)} & \cos 2 \theta_0^{(I)} & -\cos \omega_I t \sin 2 \theta_0^{(I)} \\
 \sin \omega t \sin2 \theta_0^{(I)} & 0 & \cos \omega_I t \sin 2 \theta_0^{(I)} & \cos 2 \theta_0^{(I)} \\
\end{array}
\right)\,.
\ea
Straightforward computation shows that each of the relative covariant metrics $\D_I$ has a quadruple of eigenvalues
\ba
\left(e^{ i 2\theta_0^{(I)}},e^{ i 2\theta_0^{(I)}},e^{- i 2\theta_0^{(I)}},e^{- i 2\theta_0^{(I))}}\right)
\ea
Using the $F_2$ cost function, this implies that we can compute the circuit complexity as (detailed analysis we refer to \cite{Hackl:2018ptj})
\ba\label{f2}
\math{C}_2(\bra{\W_0}\to\bra{\text{TFD}(t)})=\frac{\sqrt{\text{Tr}\lf((i\log \D)^2\rt)}}{2}=\sqrt{\sum_I |2\q_0^{(I)} |^2}\,.
\ea
If we choose generators $\hat{K}_I$ in the circuit that agree with the two-mode squeezing operations that generate $\theta_0^{(I)}$, them can be normalized by satisfying the condition
\ba
Y^I=2\theta_0^{(I)}.
\ea
 This  choice of $\hat{K}_I$ can always apply to the pairs consisting of reference and target state because $\Delta=\Omega\,\omega_0$ and can be given by
\ba
\hat{K}_I=\frac{\log \Delta_I}{||\log \Delta_I||}.
\ea
The complexity with the $\kappa=2$ measure is then given by
\ba\label{c2}
\math{C}_{\k=2}(\bra{\W_0}\to\bra{\text{TFD}(t)})=\sum_I |2\q_I|^2\,,
\ea
while in this basis, the $\kappa=1$ measure yields
\ba\label{c1}
\math{C}_{\k=1}(\bra{\W_0}\to\bra{\text{TFD}(t)})=\sum_I |2\q_I|\,.
\ea

    Note that the results of \eq{f2} and \eq{c2} are simply related, $i.e.$, $\math{C}^2_{2}=\math{C}_{\k=2}$. Thus, we know that the $F_{2}$ measure yields the same optimal trajectories with $F_{\kappa=2}$ with a test particle action in the corresponding geometry. Since that, we mainly focus on study the $\k=2$ and $\k=1$ measures in the following.
  It is also worth noting that this circuit complexity is time-independent although here we focus on the time-dependent TFD state. This result is totally different with the holographic complexity, but it is not surprising since what we considered here is a free Fermionic system while  the dual boundary conformal field theory is a strongly coupled system.

  $Remark.$--From a geometry perspective, it is understandable that the circuit complexity from the vacuum state to the time-dependent TFD state is independent of time. As explicitely demonstrated in the appendix A of \cite{Hackl:2018ptj}. the choice of the reference state $\bra{\Omega_R}$ equips the Lie group SO(4$N$) with a fibre bundle structure. That is, SO(4$N$) becomes a fibre bundle where the fibers corresponding to the different equivalence classes diffeomorphic to U(2$N$) and the base manifold identified with the space of Gaussian states is given by the quotient
  \ba
  \math{M}=\text{SO}(4N)/\text{U}(2N).
  \ea
we foliate the symplectic group by generalized cylinders defined as
\ba
C_{\sigma}=\{e^Au|\text{A}\in\text{asym}(2N),||\text{A}||=\sigma, u\in U(2N)\}
\ea
with the topology $S^{2N(2N-1)-1}\times\text{U}(2N)$. The trajectory \eq{trajectory} in the base manifold is given by
\ba
\text{U}(t)=\otimes_I e^{\theta_0^I \hat{K}_I(t)}u=\otimes_I e^{\theta_0^I (\cos(\omega_I t)A+\sin(\omega_I t)B)}u
\ea
where $A=Q_LQ_R-P_LP_R$ and $B=-(P_LQ_R+Q_LP_R)$. It is easy to check that $A,B\in \text{asym}(2N)$ and the generator $B$ is orthogonal to $A$ with $||A||=||B||$. As a result, the trajectory of the circuit lies in the generalized cylinder $\text{C}_{\sigma}$. Hence we know that the circuit complexity is time-independent.

\subsubsection{Complexity of formation}

   In this subsection, we generalize above results to the continuum case to evaluate the complexity of formation. In the above, we restrict our attention on the fermionic TFD case with 2N degrees of freedom. Taking $N\to\infty$ then leads to the continuum limit of a fermionic field theory. We consider a fermionic field theory in $n$-dimensional spacetime with mass $m$.  In that case, the label $I$ in the equations \eq{c2} and \eq{c1} will be continuous by referring, for instance, to the momentum $k$, and the summation will become the integration, $i.e.$,
   \ba
   \sum_I\to \frac{c_f V}{(2\p)^n}\int d^nk,
   \ea
   where $c_f$ denotes the internal degrees of freedom for the Fermionic field. With those in hand, the circuit complexity can be written as
\ba\label{Ck}
\math{C}_\k\lf(\bra{\Omega_0}\to \bra{\text{TFD}(t)}\rt)=\frac{c_fV}{(2\p)^{n-1}}\int d^{n-1}k |2 \arctan e^{-\b \w_k /2}|^\k \, ,
\ea
where the frequency $\w_k=\sqrt{k^2+m^2}$. One can see that we put the whole system into a box with volume $V$ to regularize the IR divergence. As for UV divergence, we normally would introduce a momentum cutoff $\Lambda$, however, in this case, one should notice that this integral is finite for the measure $\kappa=1$ and $\kappa=2$ even without the momentum cutoff $\Lambda$.

In the following, we evaluate the complexity of formation. That is, we evaluate the extra complexity required to prepare the two copies of the fermionic field theory in the TFD state compared to simply preparing each of the copies in the vacuum state with the $\kappa$ cost functions at time $t=0$, $i.e.$,
\ba
\D\math{C}_{\kappa}(\bra{\text{TFD}},\bra{\Omega_0})=\math{C}_{\kappa}(\bra{\psi_R}\to\bra{\text{TFD}})-\math{C}_{\kappa}(\bra{\psi_R}\to\bra{\Omega_0})\,.
\ea
 In the previous sections we choose the vacuum state $\bra{\Omega_0}$ as the reference state, $i.e.$, $\bra{\psi_R}=\bra{\Omega_0}$. In that case, the complexity of formation is simply equal to the complexity of the fermionic TFD state, $i.e.$
\ba
\D \math{C}_{\kappa}(\bra{\text{TFD}},\bra{\Omega_0})=\math{C}_{\kappa}(\bra{\Omega_0}\to\bra{\text{TFD}}).
\ea
This expression is exactly what described by the equation \eq{Ck}.
 The holographic complexity of formation for planar theory in $n\geq3$ is directly proportional to the entropy in both CA and CV conjectures \cite{A9}. Hence, it is natural to normalize the complexity of formation by the entropy. we consider the fermionic thermodynamic system in the subsystem $L/R$. Its partition function can be written as
\ba
\ln Z=\frac{2c_f V\W_{n-2}}{(2\p)^{n-1}}\int_0^\inf dk k^{n-2} \ln\lf(1+e^{-\b\w_k}\rt)\,.
\ea
Then, the thermodynamic entropy is given by
\ba\label{then}
S_\text{th}=\frac{\pd}{\pd T}\lf(T\ln Z\rt)=\frac{2c_f V\W_{n-2}}{(2\p)^{n-1}}\int_0^\inf dk k^{n-2}\lf[\frac{\b\w_k}{1+e^{\b \w_k}}+ \ln\lf(1+e^{-\b\w_k}\rt)\rt]
\ea
The ratio of the complexity of formation and entropy is then simply the ratio of the two functions of $\beta m$ in equations \eq{Ck} and \eq{then}. As for the massless case, the result is slightly different and it is given by
\ba\label{coes}
\frac{\D \math{C}_\k}{S_\text{th}}=\frac{A_n}{B_n}\,,
\ea
where
\ba\begin{aligned}
A_n&=\int_0^\inf du u^{n-2} |2 \arctan e^{-u/2}|^\k\,,\\
B_n&=\int_0^\inf du u^{n-2}\lf[\frac{u}{1+e^u}+ \ln\lf(1+e^{-u}\rt)\rt]\,.
\end{aligned}\ea

\begin{figure}
\centering
\includegraphics[width=3in,height=2in]{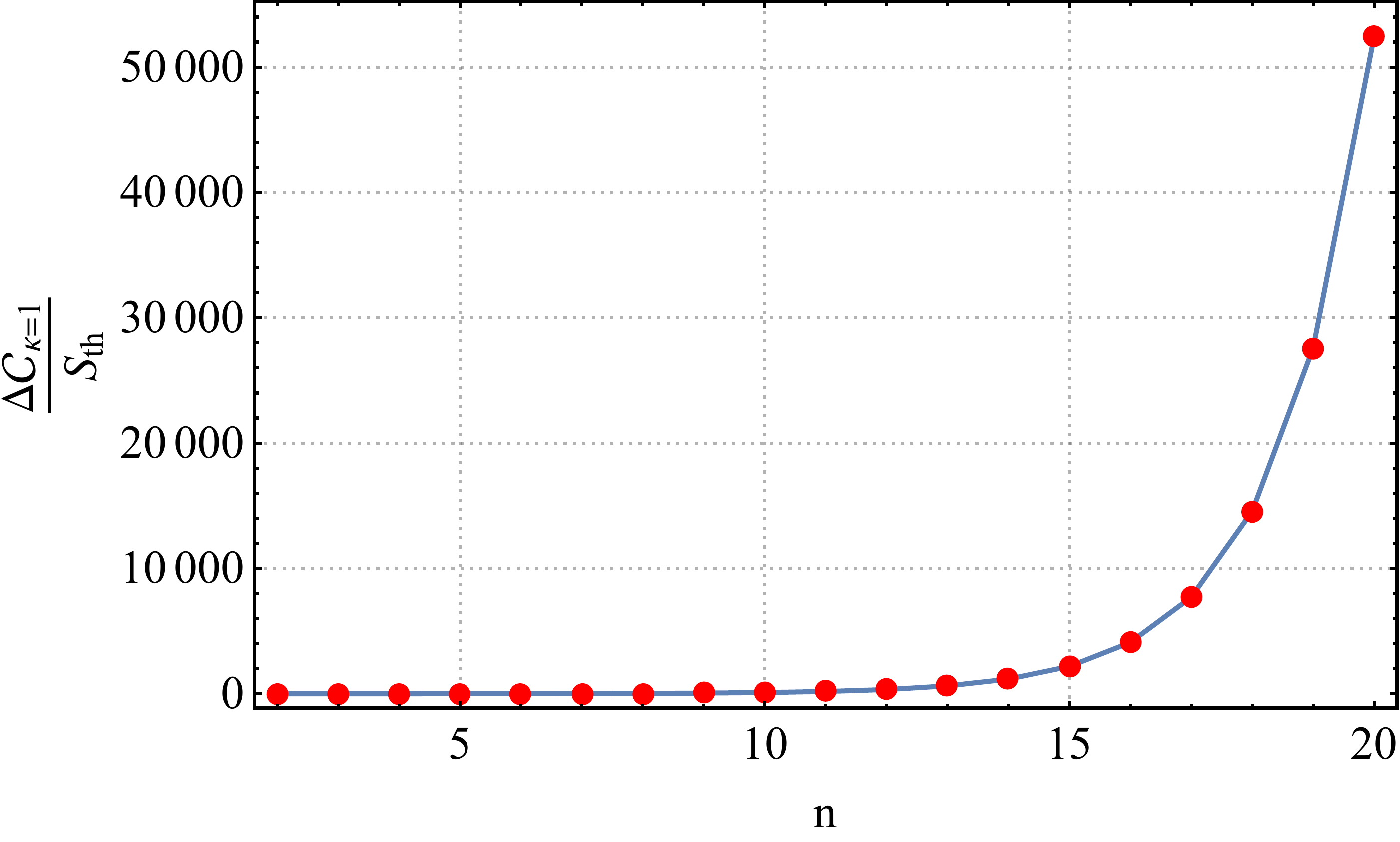}
\includegraphics[width=3in,height=2in]{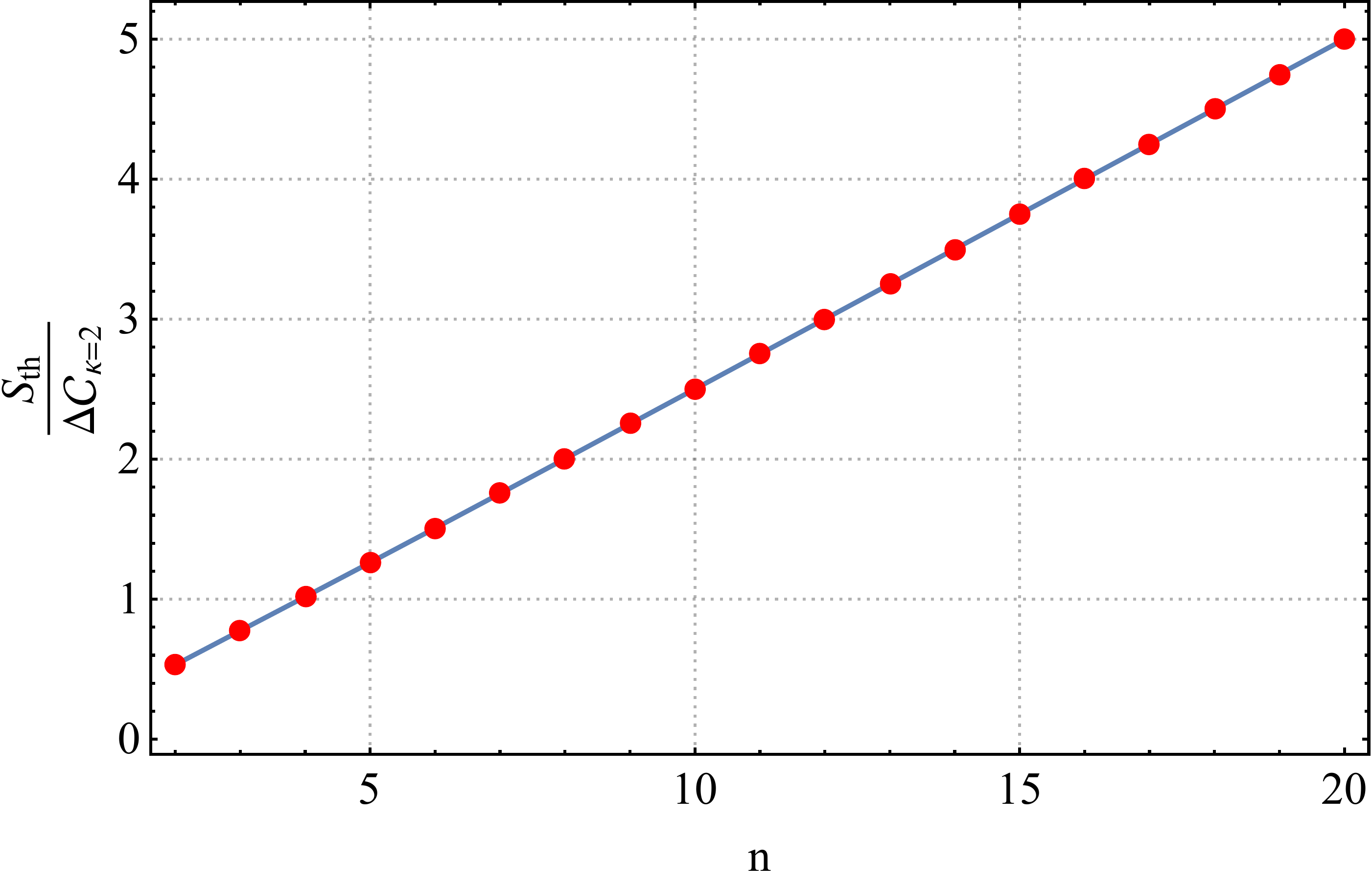}
\caption{The ration of complexity of formation and entropy in Eq. \eq{coes} $\D \math{C_\k}/S_\text{th}$, plotted as a function of the boundary dimension $n$. }
\label{CSn}
\end{figure}

Now we can numerically evaluate how the ratio of the complexity of formation and entropy behaves in a massive or massless case. And one will find our results in fermionic QFT case is extremely similar to the results for bosonic case in the section 5 of \cite{Chapman:2018hou}. We highly recommend reader to compare those results.

We show in \fig{CSn} the numerical evaluation of how the coefficient of the complexity of formation varies with the dimensions, once the theory is massless.
In the left panel of \fig{CSn}, we can see that the ration grows relatively stable when the dimension is under 15, and then grows exponentially when the boundary dimension $n$ greater than 15, for the measure $\kappa=1$. This behavior is very similar to holography, where the complexity of formation scales like the entropy, with a dimensionless coefficient that increases with the dimension of spacetime. In summary, in both our result and the holographic results, the complexity of formation is UV finite, positive and proportional to $S_\text{th}$ to leading order. However, in the right panel of \fig{CSn}, the ratio of the complexity of formation and entropy is inversely proportional to the dimension $n$ with $\kappa=2$, while in contrast the result for $\kappa=1$.

\begin{figure}
\centering
\includegraphics[width=3in,height=2in]{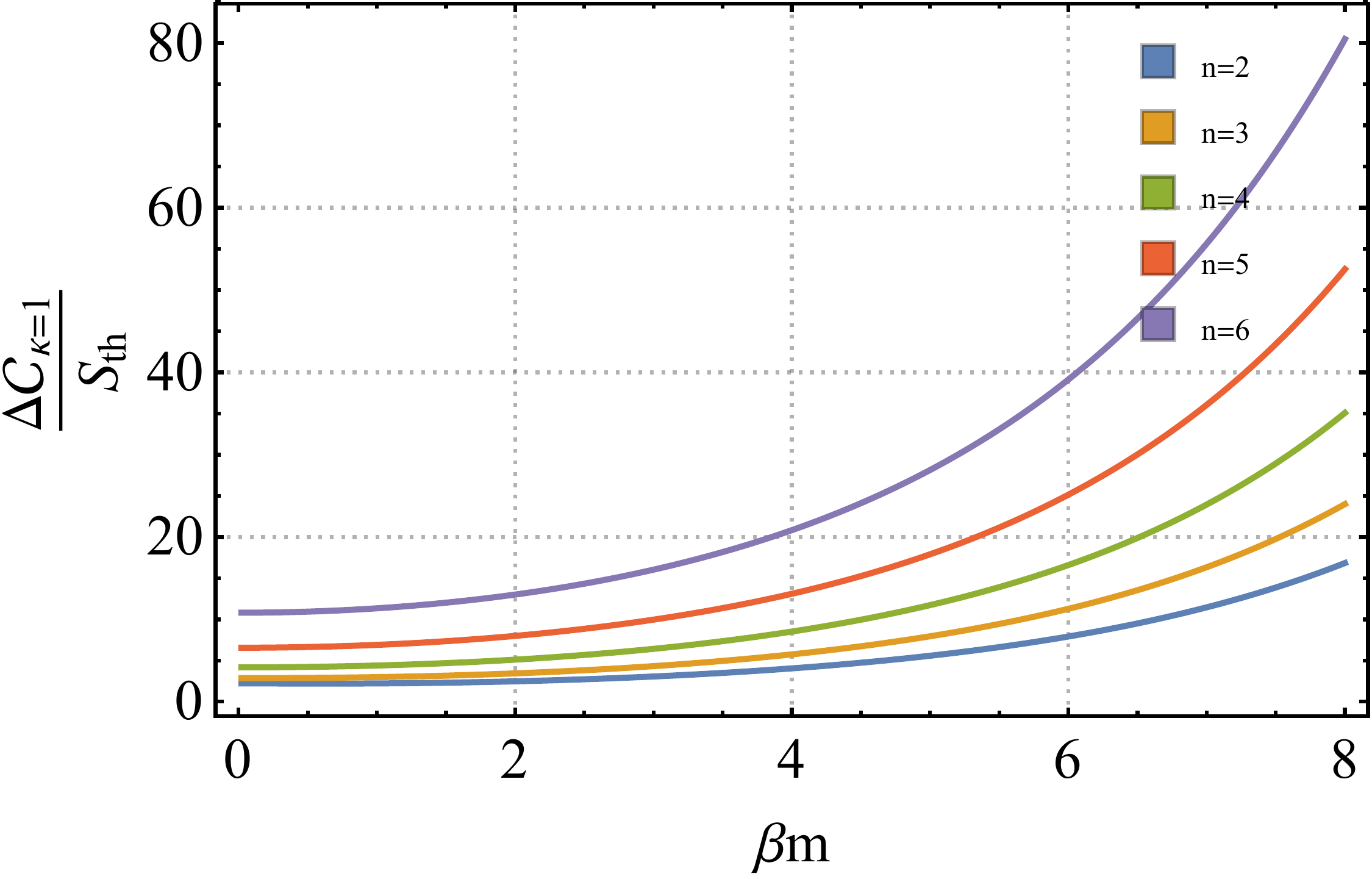}
\includegraphics[width=3in,height=2in]{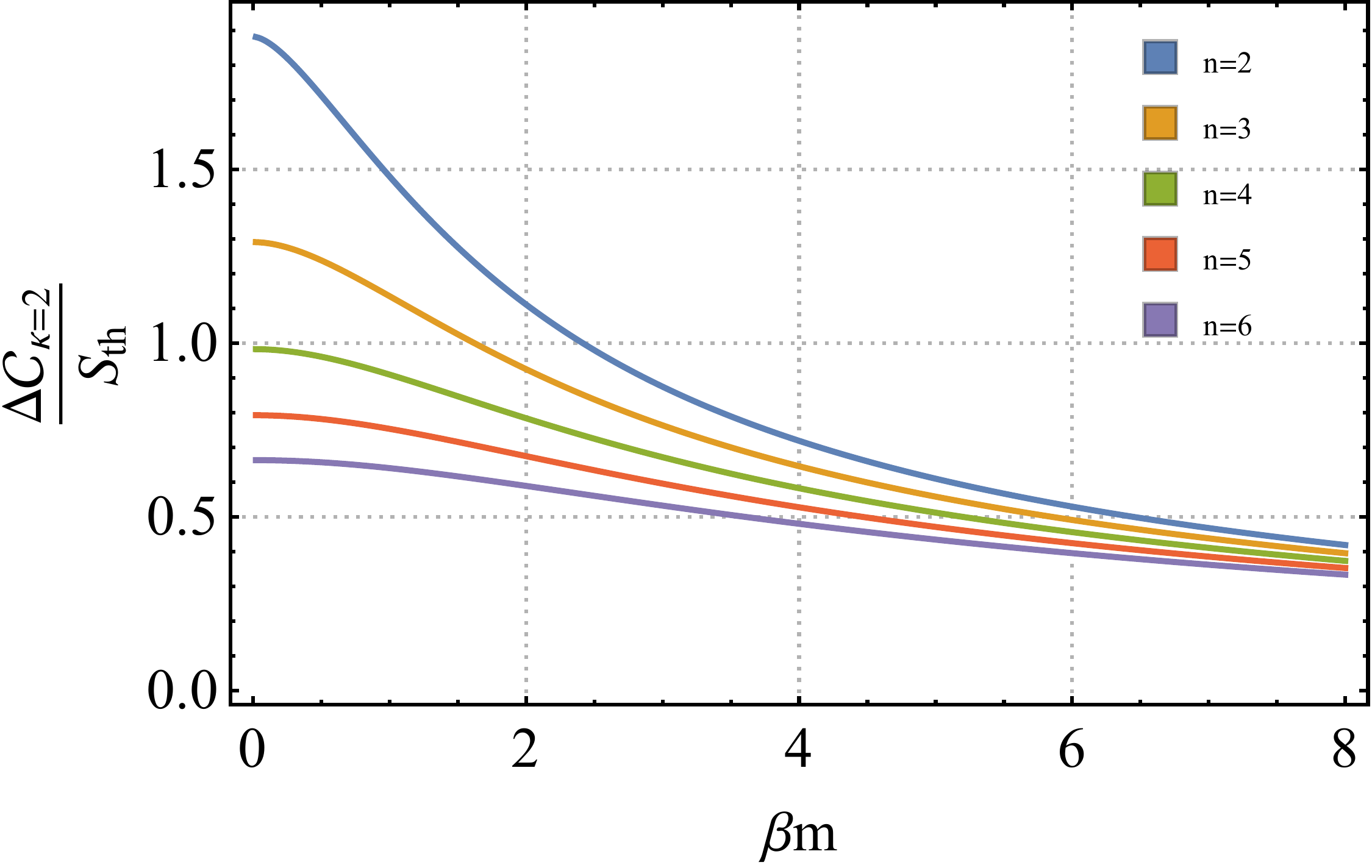}
\caption{The ration of complexity of formation and entropy in Eq. \eq{coes} $\D \math{C_\k}/S_\text{th}$ for massive case, plotted as a function of the inverse of the temperature $\b m$. }
\label{CSb}
\end{figure}
For the massive case, the ratio of the complexity of formation and entropy can be expressed as a function in terms of $\b\w$. In \fig{CSb}, the ratio of the complexity and entropy increase with respected to $\b m$ for the $\k=1$ measure while it decrease for the $\k=2$ measure.

\subsection{Circuit complexity of TFD states in free Dirac field theory}\label{CCTFD}
In the previous section, we evaluate the circuit complexity for harmonic oscillators from the vacuum reference state, which has no entanglement for the left and right subsystem, to the time-dependent target state $\bra{\text{TFD}(t)}$. In terms of this reference state, we found that the circuit complexity is time independent. It is not hard to find that, for the free Dirac field, the zero temperature reference state $\bra{\W_0}$ has the spatial entanglement. It is worth our attention to investigate the reference state as the spatial unentangled state. Therefore, in the following, we are applying Nielsen's approach tp build the optimal unitary circuit $U$, which accomplishes the transformation $\bra{\psi_T}=U\bra{\psi_R}$. The target state will be the fermionic TFD state of the Dirac field, $\bra{\psi_T}=\bra{\text{TFD}(t)}$. As reference state, we will choose a state where the local fermionic degrees of freedom at each spatial point on a given time slice are unentangled, $\bra{\psi_R}=\bra{\bar{0}}$

For simplicity, here we only consider a free Dirac field in $2$-dimensional Minkowski spacetime. The action can be written as
\ba
I=\int d^2x\bar{\y}\lf(i\g^\m\pd_\m-m\rt)\y\,.
\ea
The equation of motion can be shown as
\ba
\lf(i\g^\m\pd_\m-m\rt)\y=0\,,
\ea
with $\mu=0,1$, where
\ba
\g^0=\left(
\begin{array}{cccccccc}
 0 & 1\\
 1 & 0 \\
\end{array}
\right)\ \ \ \ \text{and}\ \ \ \ \g^1=\left(
\begin{array}{cccccccc}
 0 & 1\\
 -1 & 0 \\
\end{array}
\right).
\ea
The Dirac spinor field (on a fixed time slice, $e.g.,t=0$) can be expressed as
\ba\label{expand1}
\y(x)=\int\frac{dk}{\sqrt{2\p}}\lf(a_k u(k)e^{i k\.x}+b_k^{\dag} v(k)e^{-i k\.x}\rt)\,,
\ea
in which
\ba\begin{aligned}
u(k)=\frac{1}{\sqrt{2\w_k(\w_k-k)}}\left(
\begin{array}{cccccccc}
 \w_k-k \\
 m \\
\end{array}
\right)\,,\ \ \
v(k)=\frac{1}{\sqrt{2\w_k(\w_k-k)}}\left(
\begin{array}{cccccccc}
 \w_k-k \\
 -m \\
\end{array}
\right)\,.
\end{aligned}\ea
with $\w_k=\sqrt{k^2+m^2}$. Clearly, we have two fermionic degrees of freedom per spatial momentum $k$. Recall that the annihilation and creation operators satisfy the anti-commutation relations, $i.e.$,
\ba
\{a_k, a_{k'}^{\dag}\}=\{b_{k},b_{k'}^{\dag}\}=\d(k-k')\,.
\ea
Thw ground state is the fermionic Gaussian state $\bra{0}$, defined by $a_k\bra{0}=0=b_k\bra{0}$. The Hamiltonian can be recast by
\ba
\hat{H}=\int dk \, \w_k\lf(a_k^\dag a_k+b_k^\dag b_k\rt).
\ea
Again, it is simplest to work with the the Majorana modes which can be defined as
\ba\begin{aligned}\label{Majo1}
Q_{k}&=\frac{1}{\sqrt{2}}(a^{\dag}_{{k}}+a_{{k}})\,,\ \ \ \ \ \ P_{k}=\frac{i}{\sqrt{2}}(a^{\dag}_{k}-a_{k})\,,\\
Q'_{k}&=\frac{1}{\sqrt{2}}(b^{\dag}_{-k}+b_{-k})\,,\ \ \ P'_{k}=\frac{i}{\sqrt{2}}(b^{\dag}_{-k}-b_{-k})\,.
\end{aligned}\ea

As indicated above, our desired reference state, $\bra{\psi_R}=\bra{\bar{0}}$, will be a Gaussian state where the local fermionic degrees of freedom at each spatial point on a given time slice are unentangled. First, let introduce local creation and annihilation operators $(\bar{a}_k, \bar{a}_{k}^{\dag})$ and $(\bar{b}_k, \bar{b}_{k}^{\dag})$ satisfying
\ba
\{\bar{a}_k, \bar{a}_{k'}^{\dag}\}=\{\bar{b}_{k},\bar{b}_{k'}^{\dag}\}=\d(k-k')\,.
\ea
These operators are not completely defined until we make a specific choice on how to express the Dirac field in terms of these local operators as
\ba\label{expand2}
\y(x)=\int\frac{dk}{\sqrt{2\p}}\lf(\bar{a}_k u_0e^{i k\.x}+\bar{b}_k^{\dag} v_0e^{-i k\.x}\rt)\,,
\ea
where $u_0=u(0), v_0=v(0)$. Then our NSE reference state $\bra{\W_R}=\bra{\bar{0}}$ is defined by
$\bar{a}_k\bra{\bar{0}}=\bar{b}_k\bra{\bar{0}}=0$.
The Majorana modes of this reference state can be defined as
\ba\begin{aligned}\label{Majo2}
\bar{Q}_{k}&=\frac{1}{\sqrt{2}}(\bar{a}^{\dag}_{{k}}+\bar{a}_{{k}})\,,\ \ \ \ \ \ \bar{P}_{{k}}=\frac{i}{\sqrt{2}}(\bar{a}^{\dag}_{{k}}-\bar{a}_{{k}})\,,\\
\bar{Q}'_{{k}}&=\frac{1}{\sqrt{2}}(\bar{b}^{\dag}_{-{k}}+\bar{b}_{-{k}})\,,\ \ \ \bar{P}'_{{k}}=\frac{i}{\sqrt{2}}(\bar{b}^{\dag}_{-{k}}-\bar{b}_{-{k}})\,.
\end{aligned}\ea
One can easy verify that this reference state has no spatial entanglement, i.e.,
\ba
\langle\bar{0}|\y^\dag_A(x)\y_B(y)\bra{\bar{0}}=\d_{AB} \d(x-y)\,.
\ea
Now comparing Eqs. \eq{expand1} and \eq{expand2}, one can further obtain the Bogoliubov transformation which yields $(\bar{a}_k,\bar{a}_k^{\dag},\bar{b}_k,\bar{b}^{\dag}_k)\to(a_k,a_k^{\dag},b_k,b_k^{\dag})$,
\ba\begin{aligned}\label{transf}
\bar{a}_k&=[u_0^{\dag}u(k)]a_k+[u_0^{\dag}v(-k)]b_{-k}^{\dag}\,,\\
\bar{b}_{-k}^{\dag}&=[v_0^{\dag}u(k)]a_k+[v_0^{\dag}v(-k)]b_{-k}^{\dag}\,.\\
\end{aligned}\ea
By using Eqs. \eq{Majo1}, \eq{Majo2}, and \eq{transf}, one can obtain the transformation in terms of the Mojorana modes
\ba\begin{aligned}\label{tsf}
Q_k&=\frac{\w-k+m}{2\sqrt{\w(\w-k)}}\bar{Q}_k+\frac{\w-k-m}{2\sqrt{\w(\w-k)}}\bar{Q}'_k\,,\\
Q_k'&=\frac{\w+k-m}{2\sqrt{\w(\w+k)}}\bar{Q}_k+\frac{\w+k+m}{2\sqrt{\w(\w+k)}}\bar{Q}'_k\,,\\
P_k&=\frac{\w-k+m}{2\sqrt{\w(\w-k)}}\bar{P}_k-\frac{\w-k-m}{2\sqrt{\w(\w-k)}}\bar{P}'_k\,,\\
P_k'&=-\frac{\w+k-m}{2\sqrt{\w(\w+k)}}\bar{P}_k+\frac{\w+k+m}{2\sqrt{\w(\w+k)}}\bar{P}'_k\,.\\
\end{aligned}\ea
Then in terms of above transformation relations, if we choose the NSE state $\bra{\bar{0}}$  as a reference and  the ground state of Dirac spinor $\bra{0}$ as a target, this specific circuit can be expressed as
\ba\label{0to0}
\bra{0}=U(\hat{E}_k)\bra{\bar{0}}\ \ \ \ \text{with}\ \ \ \ U(\hat{E}_k)=e^{\int dk \a_k \hat{E}_k},
\ea
where
\ba\begin{aligned}
\a_k&=\arctan\lf(\frac{k}{\w+m}\rt)=\frac{1}{2}\arctan\lf(\frac{k}{m}\rt)\,,\\
\hat{E}_k&=\bar{a}_k^\dag \bar{b}_{-k}^\dag+\bar{a}_k \bar{b}_{-k}=\bar{Q}_k\bar{Q}'_k-\bar{P}_k\bar{P}'_k\,.
\end{aligned}\ea
Moreover,  we can write down the corresponding unitary action on the Mojorana Modes
\ba\label{barxx}
U(\hat{E}_k) \x^a U^{\dag}(\hat{E}_k)=\bar{\x}^a\,,
\ea
where $\x^a$ and $\bar{\x}^a$ denote the Majorana modes corresponding to the state $\bra{0}$ and $\bra{\bar{0}}$ respectively.

It is obvious now that the unitary operator $U(\hat{E}_k)$ in this circuit can be interpreted as an operator which entangles the spatial unentangled state. And it is also noticeable that the transformation parameter of this unitary operator $U(\hat{E}_k)$ is not convergence with respect to the momentum $k$, which implies that the cost function of the circuit between the vacuum state $\bra{0}$ and the reference state $\bra{\bar{0}}$ will be UV divergence.

Next, we turn attention to the circuit from the NSE reference state, $\bra{\psi_R}=\bra{\bar{0}}$, to the target TFD state in the Dirac system, $\bra{\psi_T}=\bra{\text{TFD}(t)}$. In the previous section, we computed the complexity for arbitrary fermionic Gaussian states for the each side of the system with a finite number of degrees of freedom, namely $N$, and the key lesson was the computations simply for two pure Gaussian state in terms of the Mojorana mode. In analogy to previous analysis, we pick up a mode $k$ in the momentum space and then consider its continuum limits. We denote
\ba\begin{aligned}
\x_k^a &=\lf(Q_{Lk},Q_{Rk},P_{Lk},P_{Rk},Q'_{Lk},Q'_{Rk},P'_{Lk},P'_{Rk}\rt), \\
\bar{\x}_k^a&=\lf(\bar{Q}_{Lk},\bar{Q}_{Rk},\bar{P}_{Lk},\bar{P}_{Rk},\bar{Q}'_{Lk},\bar{Q}'_{Rk},\bar{P}'_{Lk},\bar{P}'_{Rk}\rt).
\end{aligned}
\ea
Then, the covariance matrix of the reference state can be shown as
\ba\begin{aligned}\label{UWabU}
\W_R^{ab}&=-i\langle\W_R|\lf[\x_k^{a},\x_k^{b}\rt]\bra{\W_R}=-iM^a{}_c\langle\W_R|\lf[\bar{\x}_k^{c},\bar{\x}_k^{d}\rt]\bra{\W_R}M^b{}_d
=\lf(M\W_0M^T\rt)^{ab}\,,
\end{aligned}\ea
where $M$ denotes the inverse transformation that map $\bar{\x}_k$ to $\x_k$,$i.e.$, $\x^a=M^a{}_b\,\bar{\x}^b$, and can be further obtained from \eq{tsf}. Therefore, the covariance matrix of the reference state, $\bra{\Omega_R}=\bra{\bar{0}}$, can be shown as
\ba
\W_R=\oplus_k\W_R^{(k)}\,
\ea
with
\ba
\W_R^{(k)}=
\left(
\begin{array}{cccccccc}
 0 & 0 & \frac{m}{\omega_k } & 0 & 0 & 0 & -\frac{k}{\omega_k } & 0 \\
 0 & 0 & 0 & \frac{m}{\omega_k } & 0 & 0 & 0 & -\frac{k}{\omega_k } \\
 -\frac{m}{\omega_k } & 0 & 0 & 0 & -\frac{k}{\omega_k } & 0 & 0 & 0 \\
 0 & -\frac{m}{\omega_k } & 0 & 0 & 0 & -\frac{k}{\omega_k } & 0 & 0 \\
 0 & 0 & \frac{k}{\omega_k } & 0 & 0 & 0 & \frac{m}{\omega_k } & 0 \\
 0 & 0 & 0 & \frac{k}{\omega_k } & 0 & 0 & 0 & \frac{m}{\omega_k } \\
 \frac{k}{\omega_k } & 0 & 0 & 0 & -\frac{m}{\omega_k } & 0 & 0 & 0 \\
 0 & \frac{k}{\omega_k } & 0 & 0 & 0 & -\frac{m}{\omega_k } & 0 & 0 \\
\end{array}
\right)\,.
\ea
Following the analysis in the previous section, the covariance matrix of the target TFD state is given by
\ba
\W_T=\oplus_k\lf[\W_T^{(k)}\oplus\W_T^{(k)}\rt]\,,
\ea
where $\W_T^{(k)}$ is exactly the matrix described in \eq{WTq}. Then the relative covariance matrix of this circuit can be expressed as
\ba
\D=\oplus_k \D(k)\,
\ea
with
\ba
\D(k)=
\left(
\begin{array}{cccccccc}
 \frac{m c_{\theta }}{\omega _k} & \frac{m c_{\omega } s_{\theta }}{\omega _k} & 0 & -\frac{m s_{\theta } s_{\omega }}{\omega _k} & \frac{k c_{\theta }}{\omega _k} & \frac{k c_{\omega } s_{\theta }}{\omega _k} & 0 & \frac{k s_{\theta } s_{\omega }}{\omega _k} \\
 -\frac{m c_{\omega } s_{\theta }}{\omega _k} & \frac{m c_{\theta }}{\omega _k} & \frac{m s_{\theta } s_{\omega }}{\omega _k} & 0 & -\frac{k c_{\omega } s_{\theta }}{\omega _k} & \frac{k c_{\theta }}{\omega _k} & -\frac{k s_{\theta } s_{\omega }}{\omega _k} & 0 \\
 0 & -\frac{m s_{\theta } s_{\omega }}{\omega _k} & \frac{m c_{\theta }}{\omega _k} & -\frac{m c_{\omega } s_{\theta }}{\omega _k} & 0 & -\frac{k s_{\theta } s_{\omega }}{\omega _k} & -\frac{k c_{\theta }}{\omega _k} & \frac{k c_{\omega } s_{\theta }}{\omega _k} \\
 \frac{m s_{\theta } s_{\omega }}{\omega _k} & 0 & \frac{m c_{\omega } s_{\theta }}{\omega _k} & \frac{m c_{\theta }}{\omega _k} & \frac{k s_{\theta } s_{\omega }}{\omega _k} & 0 & -\frac{k c_{\omega } s_{\theta }}{\omega _k} & -\frac{k c_{\theta }}{\omega _k} \\
 -\frac{k c_{\theta }}{\omega _k} & -\frac{m c_{\omega } s_{\theta }}{\omega _k} & 0 & -\frac{k s_{\theta } s_{\omega }}{\omega _k} & \frac{m c_{\theta }}{\omega _k} & \frac{m c_{\omega } s_{\theta }}{\omega _k} & 0 & -\frac{m s_{\theta } s_{\omega }}{\omega _k} \\
 \frac{k c_{\omega } s_{\theta }}{\omega _k} & -\frac{k c_{\theta }}{\omega _k} & \frac{k s_{\theta } s_{\omega }}{\omega _k} & 0 & -\frac{m c_{\omega } s_{\theta }}{\omega _k} & \frac{m c_{\theta }}{\omega _k} & \frac{m s_{\theta } s_{\omega }}{\omega _k} & 0 \\
 0 & \frac{k s_{\theta } s_{\omega }}{\omega _k} & \frac{k c_{\theta }}{\omega _k} & -\frac{k c_{\omega } s_{\theta }}{\omega _k} & 0 & -\frac{m s_{\theta } s_{\omega }}{\omega _k} & \frac{m c_{\theta }}{\omega _k} & -\frac{m c_{\omega } s_{\theta }}{\omega _k} \\
 -\frac{k s_{\theta } s_{\omega }}{\omega _k} & 0 & \frac{k c_{\omega } s_{\theta }}{\omega _k} & \frac{k c_{\theta }}{\omega _k} & \frac{m s_{\theta } s_{\omega }}{\omega _k} & 0 & \frac{m c_{\omega } s_{\theta }}{\omega _k} & \frac{m c_{\theta }}{\omega _k} \\
\end{array}
\right)
\ea
in which we denote $s_\q=\sin 2\q_k, c_\q=\cos 2\q_k, s_\w=\sin\w_k t$ and $c_\w=\cos\w_k t$ with $\theta_k=\arctan(e^{-\beta \omega_k/2})$. The corresponding eigenvalues of this relative covariance matrix appear with a multiplicity of eight
and  are explicitly given by
\ba
(e^{- i \vartheta_{k_-}},e^{- i \vartheta_{k_-}},e^{- i \vartheta_{k_-}},e^{- i \vartheta_{k_-}},e^{+ i \vartheta_{k_+}},e^{+ i \vartheta_{k_+}},e^{+ i \vartheta_{k_+}},e^{+ i \vartheta_{k_+}})
\ea
with
\ba
e^{\pm i \vartheta_{k\pm}}=\tilde{\w}_k(t)\cos\lf(\tilde{\a}_k(t)\pm 2\q_k\rt)\pm i\sqrt{1-\tilde{\w}_k^2(t)\cos^2\lf(\tilde{\a}_k(t)\pm2\q_k\rt)}\,,
\ea
where we set
\ba\begin{aligned}
\tilde{\w}_k(t)&=\sqrt{(m/\w_k)^2+(k/\w_k)^2\cos^2\w_k t}\,,\ \ \
\tilde{\a}_k(t)=\frac{1}{2}\arctan\lf(\frac{k \cos \w_k t}{m}\rt)\,.
\end{aligned}\ea
$i.e.$, we have
\ba\label{thetak}
\vartheta_{k\pm}=\arctan\lf(\frac{\sqrt{1-\tilde{\w}_k^2(t)\cos^2\lf(2\a_k(t)\pm2\q_k\rt)}}{\tilde{\w}_k(t)\cos\lf(2\a_k(t)\pm 2\q_k\rt)}\rt)\,.
\ea
Then, the circuit complexity from the reference state $\bra{\psi_R}=\bra{\bar{0}}$ to the target TFD state $\bra{\psi_T}=\bra{\text{TFD}(t)}$ for the $\k=1$ and $\k=2$ measures can be directly calculated and its results are given by
\ba\label{C1C2}\begin{aligned}
\math{C}_{\k=1}(\bra{\W_R}\to\bra{\W_T})&=\frac{L}{\p}\int^{\L}_{0}dk Y_{\k=1}^{(k)}(\bra{\W_R}\to\bra{\W_T})\,,\\
\math{C}_{\k=2}(\bra{\W_R}\to\bra{\W_T})&=\frac{L}{\p}\int^{\L}_{0}dk Y_{\k=2}^{(k)}(\bra{\W_R}\to\bra{\W_T})\,,
\end{aligned}\ea
where $L$ denotes the length of a `box', $\Lambda$ denotes a momentum cutoff, and we define the per-mode complexity as
\ba\begin{aligned}
Y_{\k=1}^{(k)}(\bra{\W_R}\to\bra{\W_T})&=|\vartheta_{k+}|+|\vartheta_{k-}|\,,\\
Y_{\k=2}^{(k)}(\bra{\W_R}\to\bra{\W_T})&=\vartheta_{k+}^2+\vartheta_{k-}^2\,.
\end{aligned}\ea
Again, we introduce a box for IR divergence and a momentum cutoff for UV divergence.
The per-mode complexity $Y_{\k=1}^{(k)}\to\p/2$ and $Y_{\k=2}^{(k)}\to\p^2/8$  at the limit of large momentum $k$ for any time and hence, the corresponding complexities are UV divergent, this is why we introduced a momentum cutoff.

    Now, we consider the zero temperature limit $\b\to\inf$. Under this limit, we find
\ba
\vartheta_{k\pm}=\arctan\lf(\frac{k}{m}\rt)\,,
\ea
which is exactly the result of the Dirac vacuum state. The corresponding complexity can be expressed as
\ba\begin{aligned}\label{CC}
\math{C}_{\k=2}(\bra{\bar{0}}\to\bra{0})&=\frac{2L}{\p}\int^{\L}_{0}dk \left|\arctan\lf(\frac{k}{m}\rt)\right|^2\,,\\
\math{C}_{\k=1}(\bra{\bar{0}}\to\bra{0})&=\frac{2L}{\p}\int^{\L}_{0}dk \left|\arctan\lf(\frac{k}{m}\rt)\right|\,.
\end{aligned}\ea
Note that complexities represented by the expressions \eq{C1C2} and \eq{CC}  with respect to  measure $\kappa=1$ and $\kappa=2$, share the similar divergent behaviors, this make us wonder that it might be possible to eliminate the UV divergence by the combinations of these expressions. It is useful to define the relative complexity of the TFD state with respect to the Dirac vacuum state,
\ba
\D\math{C}_\k(\bra{\text{TFD}(t)},\bra{0})=\math{C}_{\k}(\bra{\bar{0}}\to\bra{\text{TFD}(t)})-\math{C}_{\k}(\bra{\bar{0}}\to\bra{0})\,.
\ea
At time $t=0$, it becomes the complexity of formation. And we will explore the property of convergence of this relative complexity in the following.
\subsubsection{Complexity of the TFD at $t=0$}
In this subsection, we focus on the special case of the time-independent TFD state at $t=0$ given by \eq{TFDt0}. In this situation, we found $\tilde{\w}_k=1$ and $\tilde{\a}_k=\a_k=\arctan\lf(k/m\rt)$. Then, according to \eq{thetak}, we have
\ba\label{thetak0}
|\vartheta_{k\pm}|=\left|\arctan[\tan2\lf(\q_k\pm\a_k\rt)]\right|\,.
\ea
Since $0<\q_k\leq \p/4$ and $0\leq\a_k(t)<\p/4$, one can obtain
\ba\begin{aligned}\label{thetak1}
|\vartheta_{k-}|&=2\left|\q_k-\a_k\right|\,,\\
|\vartheta_{k+}|&=
\left\{
\begin{array}{cccccccc}
 2\left(\q_k+\a_k\right)\ \ \ \ \ \ \ \ \ \q_{k}+\a_{k}\leq \p/4\\
\p-2\left(\q_k+\a_k\right) \ \ \ \ \q_{k}+\a_{k}> \p/4\\
\end{array}
\right.\,.
\end{aligned}\ea

Firstly, let us consider the massless case where $m\to0$. Then, we have $\a_k=\p/4$. According to \eq{thetak1}, we can find $|\vartheta_{k\pm}|=\p/2-2\q_k$. For the $\k=2$ measure, we have
\ba\label{0C2}\begin{aligned}
\D \math{C}_{\k=2}(\bra{\text{TFD}(t)},\bra{0})&=\frac{2L}{\p}\int^{\inf}_{0}dk\lf[ \lf(\frac{\p}{2}-2\q_k\rt)^2-\frac{\p^2}{4}\rt]\\
&=\frac{4L}{\p}\int^{\inf}_{0}dk\q_k\lf(2\q_k-\p\rt)\\
&=-16.8288\times\frac{T L}{\p}\,.
\end{aligned}\ea
For the $\k=1$ measure, with similar calculation, one can obtain
\ba\label{0C1}
\D \math{C}_{\kappa=1}(\bra{\text{TFD}(t)},\bra{0})=-\frac{4L}{\p}\int_0^\inf dk \q_k=-7.32772\times\frac{T L}{\p}\,.
\ea
These results show that the complexity of formation is in direct proportional to the temperature $T$ in the massless case. Furthermore, the relative complexity of the TFD state with respect to the Dirac vacuum state for the measure $\kappa=1$ and $\kappa=2$  share almost the same expression. Moreover, above calculation shows that the UV divergence of the circuit from $\bra{\bar{0}}$ to the TFD state can be totally eliminated by minus the circuit complexity from $\bra{\bar{0}}$ to the zero temperature state.

Next, we consider the massive case. Here let us define some useful the dimensionless variables first,
\ba\label{dmless1}
\tilde{k}:=k/m\,,\ \ \ \tilde{\b}:=m\b\,,\ \ \ \tilde{L}:=m L\,.
\ea
By virtue of the property of the Eq.\eq{thetak0}, the turning point of the inverse of temperature $\tilde{\b}_c=1.32546$ can be obtained from the equation $\q_{k}+\a_{k}=\p/4$.

\begin{figure}
\centering
\includegraphics[width=3in,height=2in]{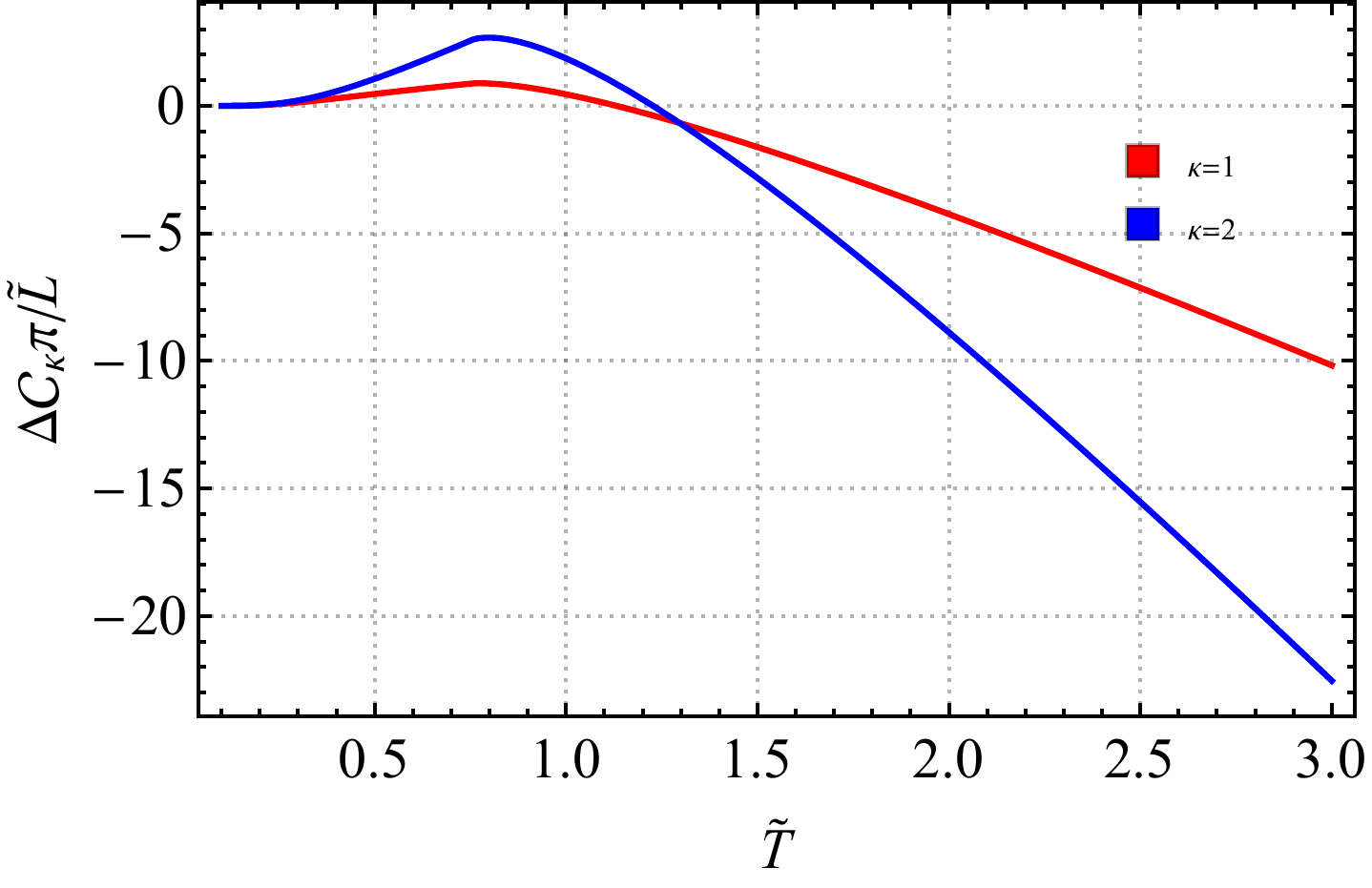}
\caption{The relative complexity for $\k=1$ (red) and $\k=2$ (blue) measures.}
\label{Ck2m}
\end{figure}

According to Eq.\eq{thetak1}, for the situation $\tilde{\b}>\tilde{\b}_c$, the circuit complexity can be written as
\ba\label{C2}
\D\math{C}_{\k=2}(\bra{\text{TFD}(t)},\bra{0})=\frac{8\tilde{L}}{\p}\int^{\inf}_{0}d\tilde{k} \lf|2\arctan e^{- \tilde{\b}\tilde{\w}_k/2}\rt|^2\,,
\ea
which shares the same formula with the circuit complexity between the TFD state and the vacuum state $\bra{0}$. We can also see that this complexity of formation is UV convergence in this situation. In \fig{Ck2m}, we present the relationship between the dimensionless temperature $\tilde{T}$ and the relative complexity of the $\k=1$ and $\k=2$ measures. At the beginning, both trajectories are positive, increase to their maximum values when both reach exactly the same turning point, after that both trajectories decrease rapidly to the $-\inf$.

\subsubsection{Complexity of the TFD at general $t$}
In this subsection, we consider the time dependent complexity for the Fermionic TFD state in the free Dirac field.

\begin{figure}
\centering
\includegraphics[width=3in,height=2in]{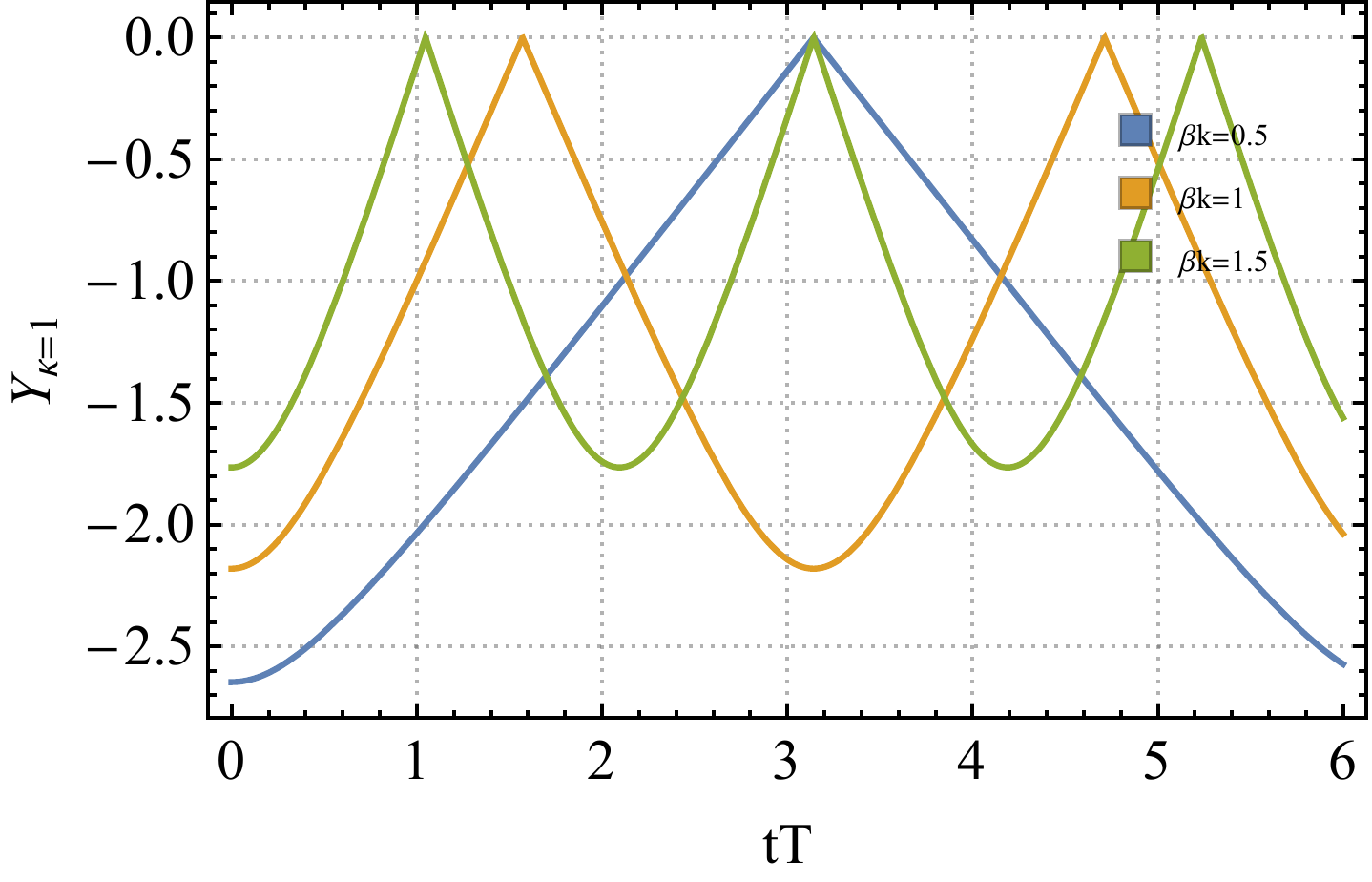}
\includegraphics[width=3in,height=2.05in]{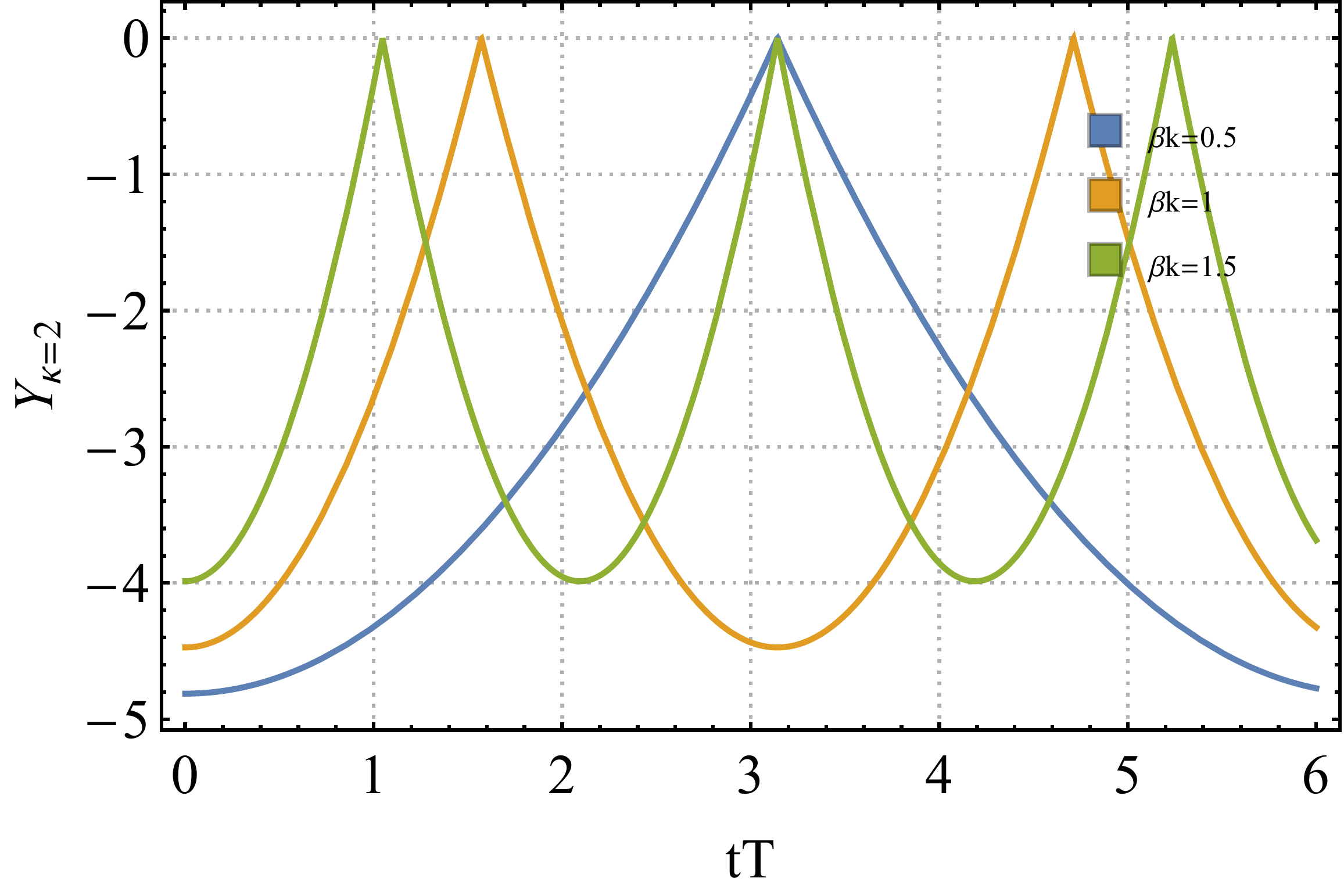}
\caption{The time involution of the per-mode complexity for $\k=1$ (left) and $\k=2$ (right) measures in the massless case.}
\label{Ykm0}
\end{figure}

\begin{figure}
\centering
\includegraphics[width=3in,height=2in]{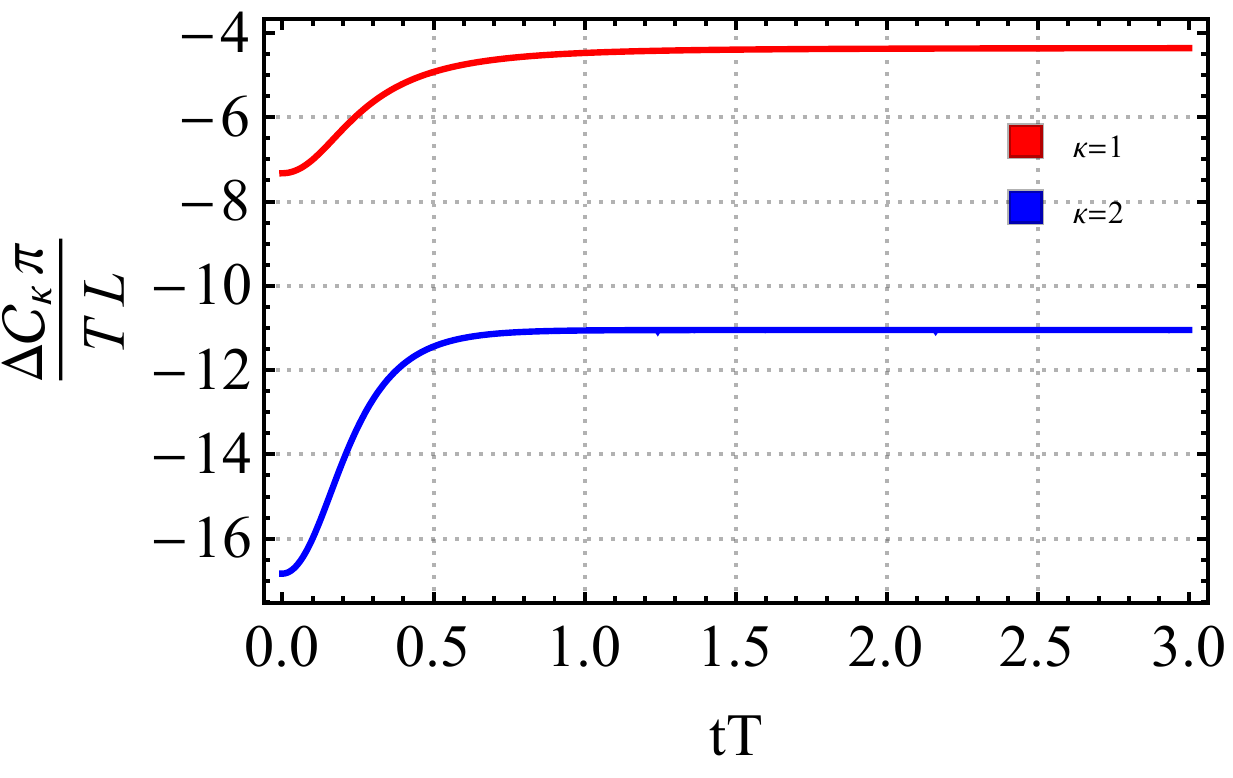}
\includegraphics[width=3in,height=2.05in]{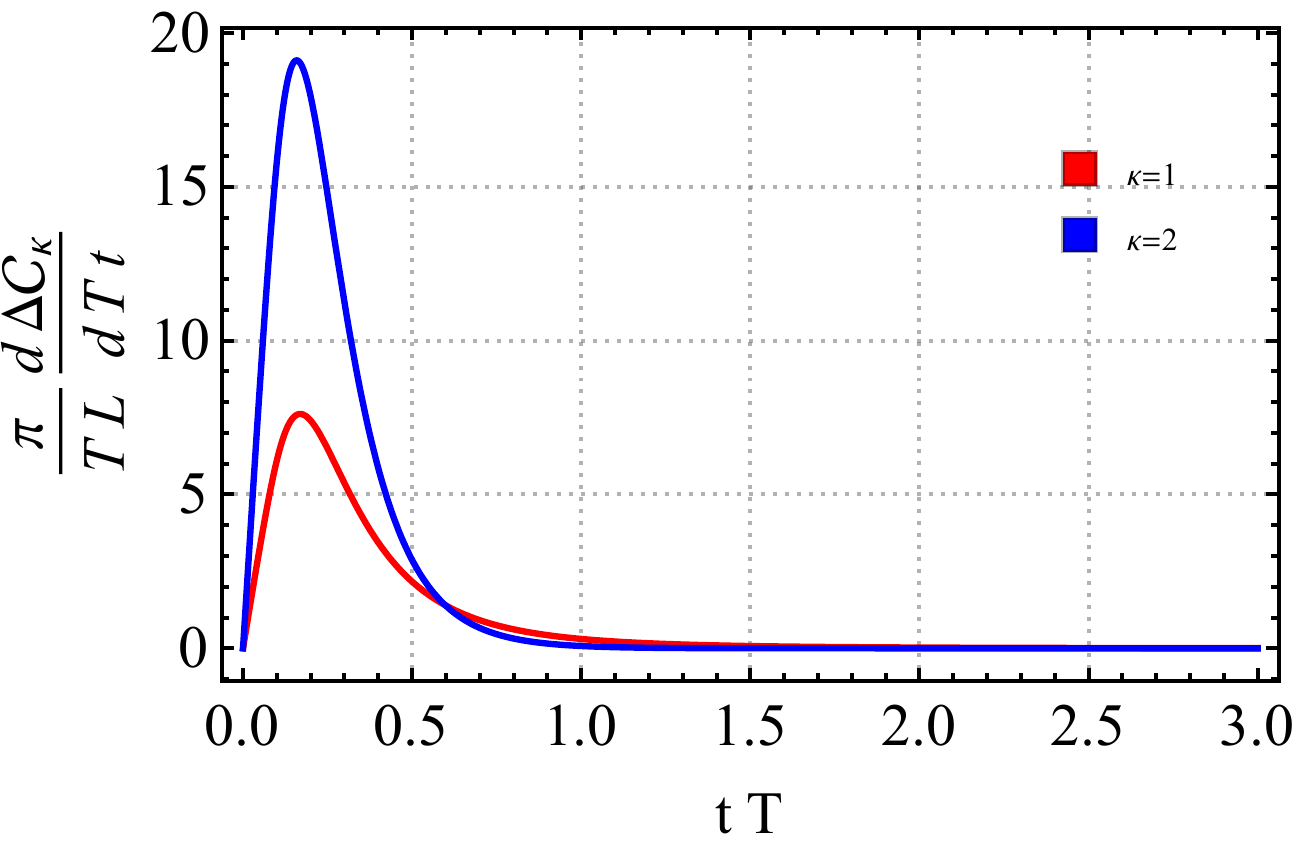}
\caption{The time involution of the complexity of formation (left) and their time derivative (right) for $\k=1$ (red) and $\k=2$ (blue) measures in the massless case.}
\label{Ckm0}
\end{figure}
For concreteness, let us focus on the massless case first. Under the massless limit $m\to 0$, one can obtain
\ba
|\vartheta_{k\pm}|=\arctan\lf(\sqrt{\cosh ^2\left(\frac{\beta  k}{2}\right) \sec ^2k t-1}\rt)\,.
\ea
Then, the per-mode complexity can be written as
\ba\begin{aligned}
Y_{\k=2}(k)&=2 \lf[\arctan\lf(\sqrt{\cosh ^2\left(\frac{\beta  k}{2}\right) \sec ^2k t-1}\rt)\rt]^2-\frac{\p^2}{2}\,,\\
Y_{\k=1}(k)&=2 \arctan\lf(\sqrt{\cosh ^2\left(\frac{\beta  k}{2}\right) \sec ^2k t-1}\rt)-\p\,.\\
\end{aligned}\ea

In \fig{Ykm0}, we plot the time dependence of per-mode complexity with the variation of the momentum $\b k$. We can see that this circuit complexity oscillates in time, and the amplitude of the oscillations decrease as $\b k$ increase, while the frequency of the oscillations increase as $\b k$ increase.

In \fig{Ckm0}, we present the time dependence of the relative complexity for the massless case in $\k=1$ and $\k=2$ measures respectively. We find that the relative complexity start with a negative value, increase with the evolve of time, then rapidly saturates at a constant which is still under zero on a time scale of order the inverse temperature. This shares a similar behaviors with the bosonic system discussed in Ref.\cite{Chapman:2018hou}. But differently, the result $\Delta\math{C_{\kappa}}(t)-\Delta\math{C_{\kappa}}(0)$ of the fermionic system is positive, while it is negative in the bonsonic system. Moreover, we see that the complexity growth rate increase at the early time and decrease later, finally goes to zero at the large time limit. This is very different from the CV and CA conjectures where the complexity is always increase at the late time. However, this result is not unexpected, since the boundary CFT is a strongly coupled system and here we investigate the free system.

\begin{figure}
\centering
\includegraphics[width=3in,height=2in]{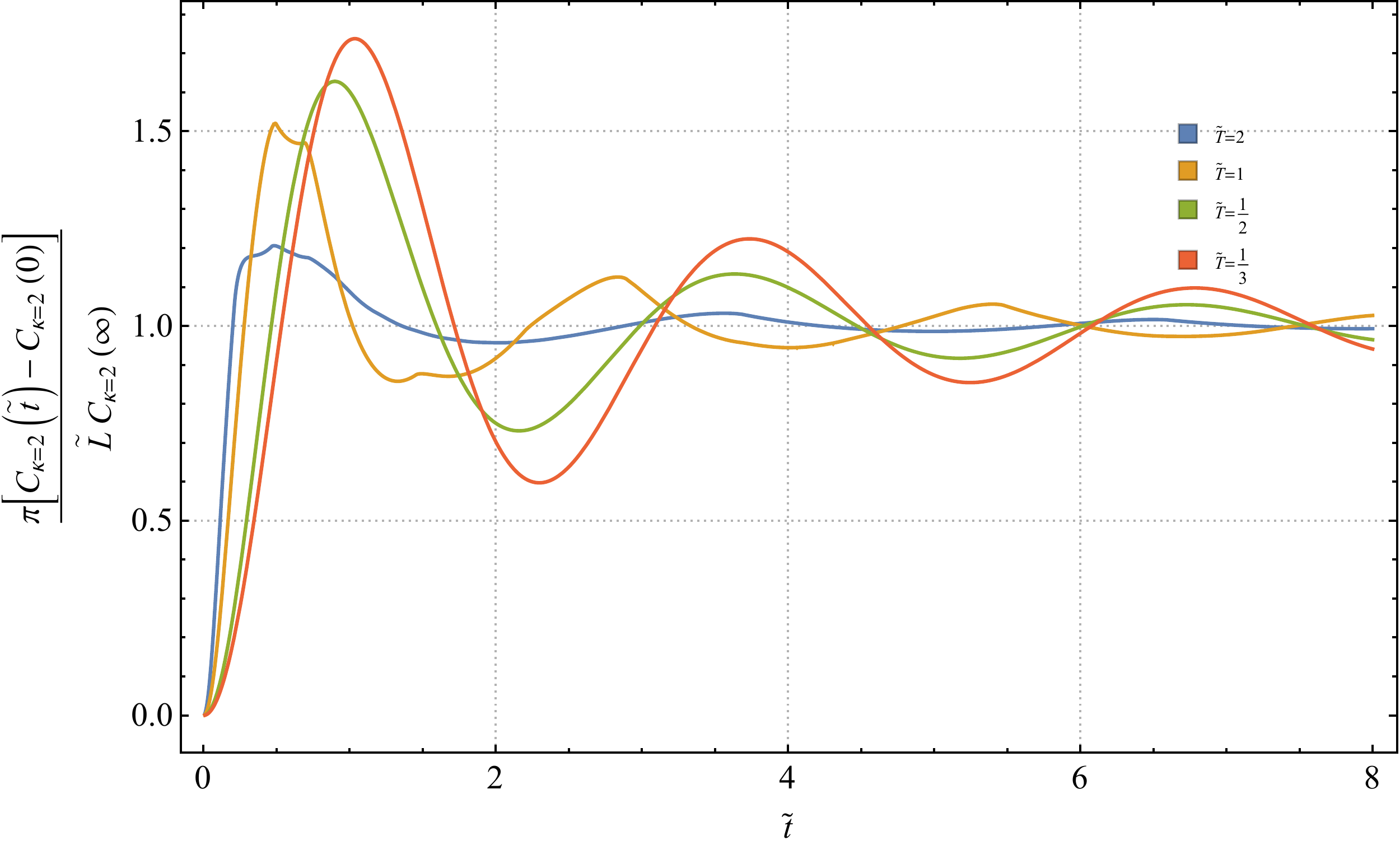}
\includegraphics[width=3in,height=2in]{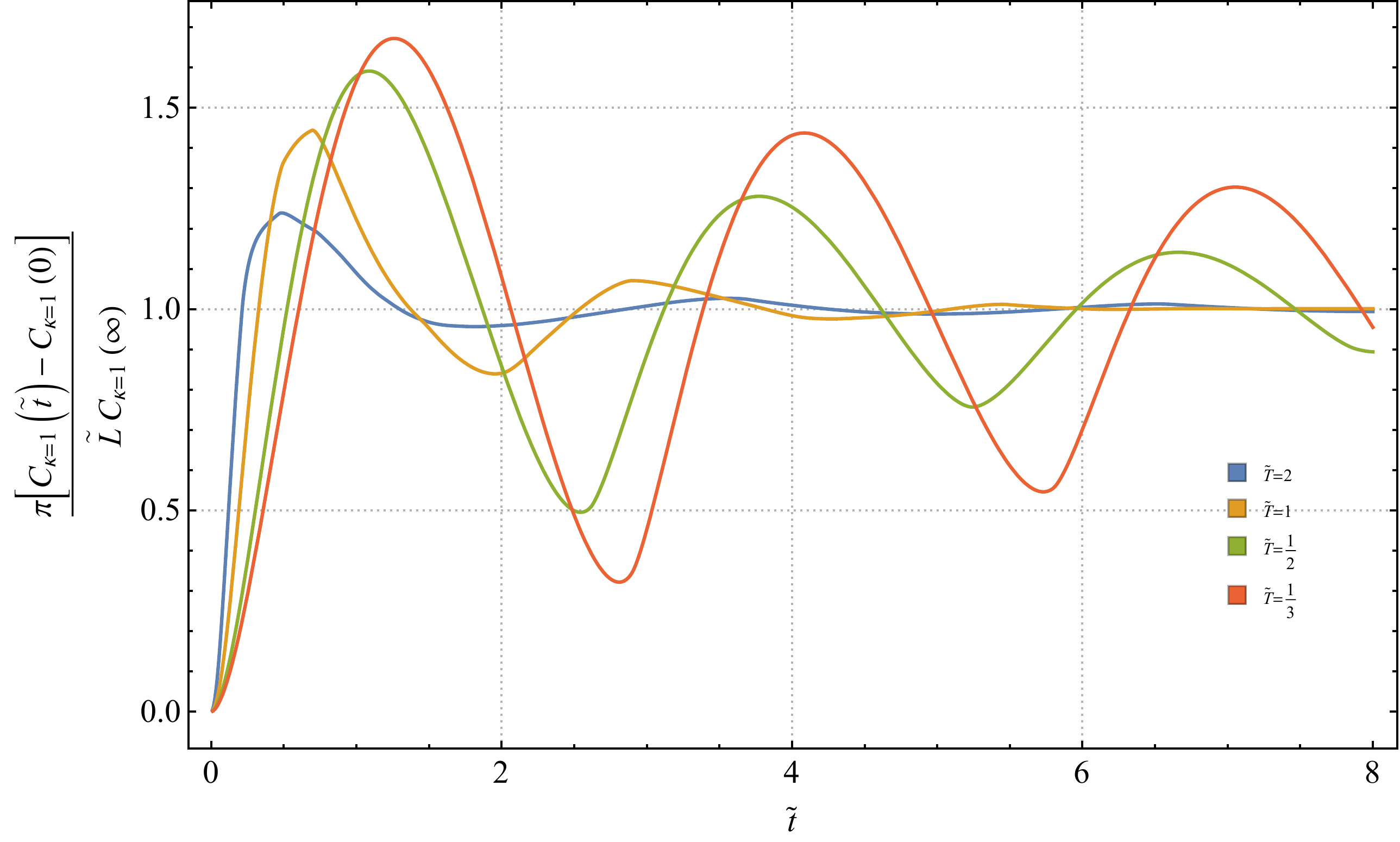}
\caption{The time evolution of the relative complexity for $\k=2$ (left) and $\k=1$ (right) measures.}
\label{Ctk2m}
\end{figure}
\begin{figure}
\centering
\includegraphics[width=3in,height=2in]{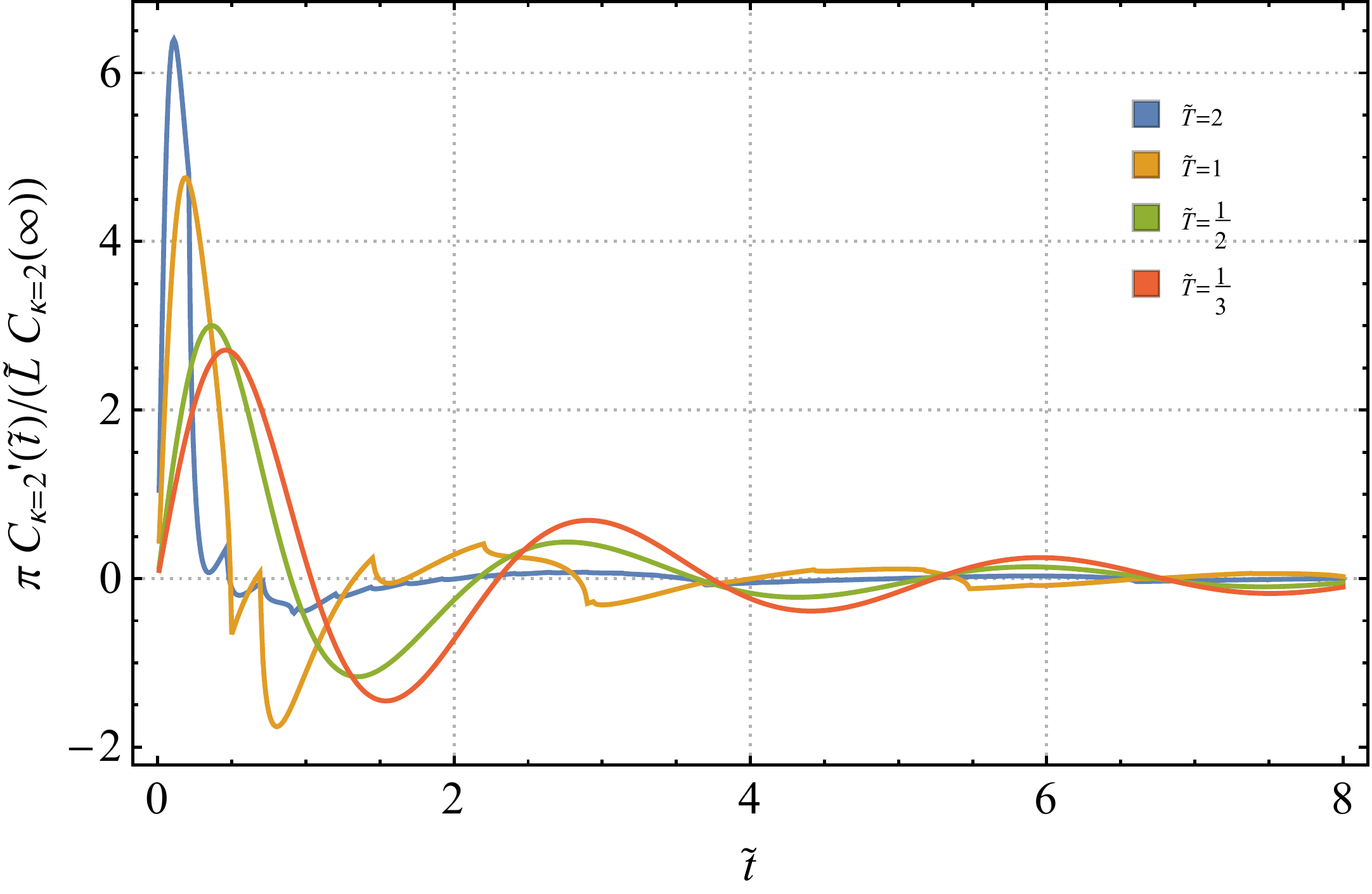}
\includegraphics[width=3in,height=2in]{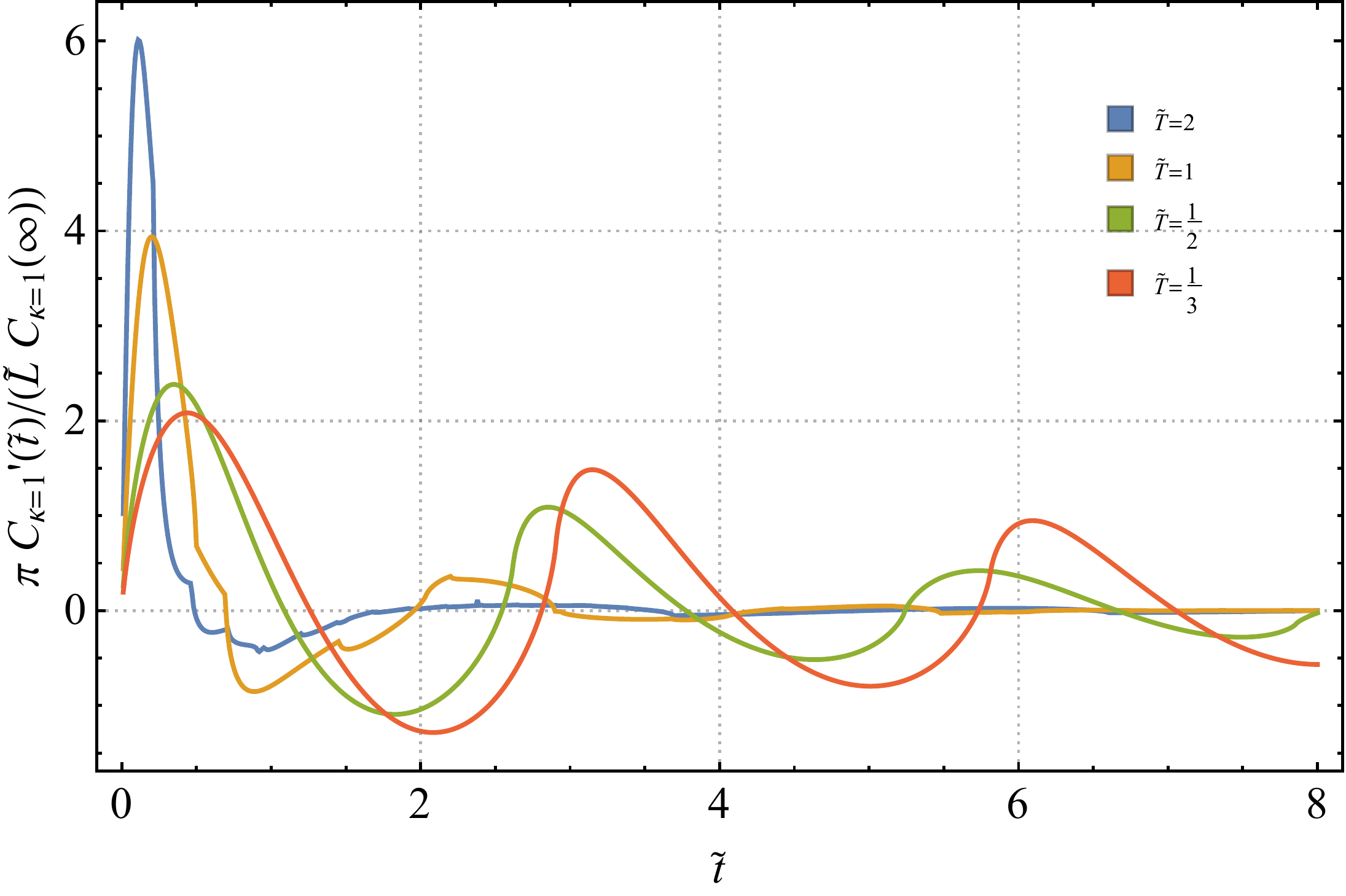}
\caption{The time derivative of the relative complexity for $\k=2$ (left) and $\k=1$ (right) measures.}
\label{dCtk2m}
\end{figure}
\begin{figure}
\centering
\includegraphics[width=4in,height=2.8in]{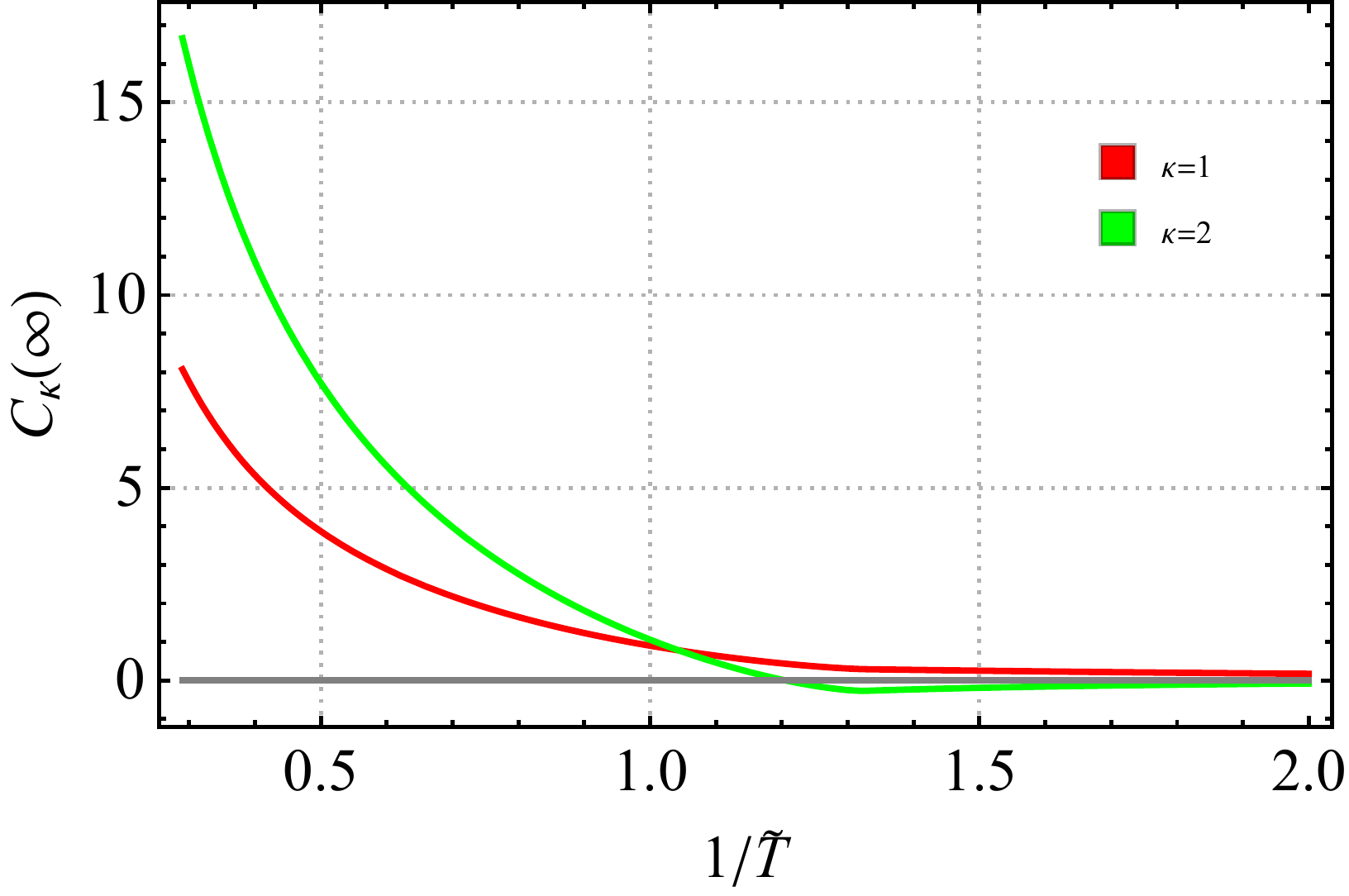}
\caption{The asymptotic value of the complexity respect with the inverse of the dimensional temperature $1/\tilde{T}$ }
\label{MM}
\end{figure}
Next, we consider the massive case of the TFD state at general $t$. Following the discussion in the massless case, here we define the dimensionless variables as \eq{dmless1}. Moreover, we also define the dimensionless time $\tilde{t}=m t$.

In \fig{Ctk2m}, we investigate the time evolution of the complexity for different values of the dimensionless temperature $\tilde{T}$. To make these figures more transparent, we consider the normalization complexity which is divided by the asymptotic values. The plots show damped oscillations, at late times we observe a saturation to a constant value. Those features matches the behaviors of the massive case of the bosonic system and the massquench case for the free Fermionic system. Further, as shown in these figures, one can find that there might exist a turning point $\tilde{\b}_c$ in the time evolution cases, where for the cases $\tilde{\b}\leq \tilde{\b}_c$ the curves are not smooth. Moreover, we can also see that there is an oscillatory behaviour with period $\D \tilde{t}\sim \p$. In \fig{dCtk2m}, we show the growth rate also has the behaviors of the damped oscillation, and finally it goes to a constant value zero. Finally, we illustrate the asymptotic value of the complexity respect with the inverse of the dimensional temperature in \fig{MM}. One can see that it increase with respected to the dimensionless temperature $\tilde{T}$.
Different with $\k=1$ measures, for the $\k=2$ measures, there exist a zero point in its figure, and finally approaches zero from below zero.

\section{Fubini-Study metric approach for circuit complexity of the TFD state}\label{FSmetric}
In this section, we apply the Fubini-Study(FS) approach proposed in \cite{Chapman:2017} to evaluate the circuit complexity of the Fermionic TFD state. This method is based on the Fubini-Study metric and the unitary transformation to define a geometry on the space of states.

 We first consider a trajectory described by a particular choice of $\l^\m(\s)$. Then, the corresponding states can be generated by
\ba\label{path}
\bra{\y(\s)}=\math{P}\exp\lf[-i\int^{\s}_{0}ds\hat{G}(s)\rt]\bra{\y_\text{R}}\,,\ \ \ \ \text{where}\ \ \hat{G}(s)=\sum_{\m}\dot{\l}^{\m}(s)\math{O}_{\m}
\ea
where $\math{O}_\m\in \math{G}$ is the set of Hermitian operators which generate the evolution of the state $\bra{\y(\s)}$. Then, according to the FS line element
\ba
ds_{\text{FS}}(\s)=d \s\sqrt{\lf|\pd_s\bra{\y(s)}\rt|^2-\lf|\langle\y(s)|\pd_s\bra{\y(s)}\rt|^2}\,,
\ea
the length of a path can be described by
\ba
\math{L}\lf(\bra{\y(s)}\rt)=\int^{s}_{0}ds_{\text{FS}}(\s)\,.
\ea
The complexity between the target state $\bra{T}$ and the reference state $\bra{R}$ is given by the minimal length of the path $\bra{\y(0)}=\bra{R}$ to $\bra{\y(\s_f)}=\bra{T}$ derived by $\hat{G}(\s)$ in $\math{G}$,
\ba
\math{C}_\text{FS}\lf(\bra{R}\to\bra{T}\rt)=\min_{\hat{G}(s)}\math{L}\lf(\bra{\y(s)}\rt)\,.
\ea
One can find that the definition of the circuit complexity is based on the choice of the generator set $\math{O}_\m$. Finding the complexity in a very general generator set seems to be a too mathematical and technical problem. Therefore, we only consider the special generator set which can generate all of our target state from our reference state. Obviously, this requirement make the complexity rely on the choice of the reference state. For finding a better method among the Fubini-Study metric approach and Nielsen's geometric method to match the result from holography, here we follow some discussion in the previous sections for the Nielsen approach,  and also consider the cases for the zero temperature reference state $\bra{\Omega_0}$ and the reference state $\bra{\bar{0}}$ which has no spatial entanglement, separately.

\subsection{Zero temperature reference state}
In this subsection, we consider the case with the zero temperature reference state. Therefore, here we only need to evaluate the evolution of the complexity between the fermionic TFD state and the zero temperature state $\bra{\W_0}$. According
to \eq{TFDt}, one can find that the unitary operator can be generated by the hermitian operator
\ba
K_1\equiv\frac{K_++K_-}{2 i}\,,\ \ \ \ \ K_2\equiv\frac{K_+-K_-}{2}\,,
\ea
with
\ba
K_+=a_L^{\dag}a_R^\dag\,,\ \ \ \ K_-=a_La_R\,.
\ea
Note that the set of generators should form a closed Lie-algebra. The minimal set can be chosen as
\ba
\math{G}=\text{span}\left\{K_0,K_1,K_2\right\}\,,
\ea
in which we defined
\ba
K_0=\frac{1}{2}\lf(a_L^\dag a_L+a_R^\dag a_R-1\rt)\,.
\ea
Then, one can verify that these generators satisfy the $su(1,1)$ Lie-algebra,
\ba
[K_-,K_+]=2 K_0\,,\ \ \ \ [K_0, K_{\pm}]=\pm K_\pm\,.
\ea
The trajectory can be expressed as
\ba
\bra{\y(\s)}=\hat{U}(\s)\bra{\W_0}=e^{\hat{g}(\s)}\bra{\W_0}
\ea
with
\ba
\hat{g}(\s)=\a^+(\s)K_++\a^-(\s)K_-+\a^0(\s)K_0\,.
\ea
Since the generator set forms a complete Lie-algebra, the unitary operator can be decomposed as
\ba
\hat{U}(\s)=e^{\g^+(\s) K_+}e^{\g^0(\s) K_0}e^{\g^-(\s) K_-}\,.
\ea
By using the identities
\ba
K_-\bra{\W_0}=0\,,\ \ \ K_0\bra{\W_0}=-\frac{1}{2}\,,
\ea
the trajectory can be recast by
\ba
\bra{\y(\s)}=e^{-\g^0(\s)/2}e^{\g^+(\s) K_+}\bra{\W_0}\,.
\ea
By the normalization, one can obtain the constrain $\g^0(\s)=\ln\lf(1+|\g^+(\s)|^2\rt)$. Then, the complexity between $\bra{TFD(t)}$ and $\bra{\W_0}$ can be obtained by
\ba\label{CFS}
\math{C}_\text{FS}\lf(\bra{\W_0}\to\bra{TFD(t)}\rt)=\min\int_{0}^{s}d\sigma\sqrt{\frac{c_fV}{(2\p)^d}\int d^dk\frac{|\pd_\s \g^+_k(\s)|^2}{(1+| \g^+_k(\s)|^2)^2}}\,.
\ea
Therefore, to calculate the complexity is equivalent to find the geodesic of the line element with the boundary condition $\g^+_k(\s_f)=e^{-\b\w_k/2-i\w_k t}$. For convenience, we define
\ba
\gamma^+_k(\s)=\tan\lf(\frac{\q_k(\s)}{2}\rt) e^{i \f_k(\s)}\,.
\ea
Then, the FS line element can be written as
\ba\label{FSmatric}
ds_{\text{FS}}^2=\frac{1}{4}\frac{c_fV}{(2\p)^{n-1}}\int_{|\bm{k}|<\L} d^{n-1}k\lf(d\q_k^2+\sin^2\q_k d\f_k^2\rt)\,.
\ea
Now, it is clear that this line element corresponds to an integral over an upper hemisphere. And the boundary conditions become
\ba\begin{aligned}
\q_k(\s_f)&=2\arctan e^{-\b\w_k/2}\,,\\
\f_k(\s_f)&=-\w_k t\,.
\end{aligned}\ea
Then, the circuit complexity can be obtained by
\ba\label{COFS}
\math{C}_\text{FS}^2\lf(\bra{\W_0}\to\bra{TFD(t)}\rt)
=\frac{c_fV}{(2\p)^{n-1}}\int d^{n-1}k\lf|\arctan e^{-\b\w_k/2}\rt|^2\,,
\ea
which has actually the similar formula \eq{Ck} with the circuit complexity from the Nielsen approach between the TFD state and the Dirac vacuum reference state. Note that in this case, the complexity is also time independent. This feature can be explained by the geometry of the FS matric \eq{FSmatric}. The reference state is located on the north pole $\q_k=0$, and the trajectory of the time evolution is located on the circle $\q_k=\q_k(s)$, which has the same length to the north pole, which will reflect on the time independent features.

\subsection{Change reference state}
In this subsection, we consider the case with the NSE state as the reference state $\bra{\W_R}=\bra{\bar{0}}$. According to Eqs. \eq{TFDt} and \eq{0to0}, one can find
\ba\begin{aligned}
\bra{TFD(t)}&=e^{z \lf(a_L^{\dag}a_R^{\dag}+b_L^{\dag}b_R^{\dag}\rt)+z^*\lf(a_La_R+b_Lb_R\rt)}e^{\a \hat{E}}\bra{\bar{0}}_L\bra{\bar{0}}_R\\
=&e^{\a \hat{E}}e^{z \lf(\bar{a}_L^{\dag}\bar{a}_R^{\dag}+\bar{b}_L^{\dag}\bar{b}_R^{\dag}\rt)+z^*\lf(\bar{a}_L\bar{a}_R+\bar{b}_L\bar{b}_R\rt)}\bra{\bar{0}}_L\bra{\bar{0}}_R\,,
\end{aligned}\ea
where we used the relation \eq{barxx}
\ba\label{barxx1}
e^{-\a \hat{E}}\x^a e^{\a_k \hat{E}}=\bar{\x}^a\,
\ea
with
\ba
\hat{E}=\bar{a}_L^{\dag}\bar{b}_L^{\dag}+\bar{a}_R^{\dag}\bar{b}_R^{\dag}+\bar{a}_L\bar{b}_L+\bar{a}_R\bar{b}_R\,.
\ea
From above expressions, one can find a nontrivial set of generators $\math{G}$  which generate transformations between the TFD state $\bra{\text{TFD}(t)}$ and the reference state $\bra{\bar{0}}$,
\ba
\math{G}=\text{span}\left\{U,V, K_1\equiv\frac{K_++K_-}{2 i},K_2\equiv\frac{K_+-K_-}{2}, E_1\equiv\frac{E_++E_-}{2 i},E_2\equiv\frac{E_+-E_-}{2}\right\}\,,\nn
\ea
with
\ba\begin{aligned}
K_+&=\bar{a}_L^{\dag}\bar{a}_R^{\dag}+\bar{b}_L^{\dag}\bar{b}_R^{\dag}\,,\ \ \ \ K_-=\bar{a}_L\bar{a}_R+\bar{b}_L\bar{b}_R\,,\\
E_+&=\bar{a}_L^{\dag}\bar{b}_L^{\dag}+\bar{a}_R^{\dag}\bar{b}_R^{\dag}\,,\ \ \ \ E_-=\bar{a}_L\bar{b}_L+\bar{a}_R\bar{b}_R\,,\\
U&=\frac{1}{2}\lf(\bar{a}_L^\dag\bar{a}_L+\bar{b}_L^\dag\bar{b}_L+\bar{a}_R^\dag\bar{a}_R+\bar{b}_R^\dag\bar{b}_R-2\rt)\,,\\
V&=\frac{1}{2}\lf(\bar{a}_L^\dag\bar{b}_R+\bar{b}_R^\dag\bar{a}_L+\bar{b}_L\bar{a}_R^\dag+\bar{a}_R\bar{b}_L^\dag\rt)\,.
\end{aligned}\ea
Then, one can further verify these generators construct a Lie-algebra. And the non-vanish Lie brackets can be shown as
\ba\begin{aligned}
\left[K_-,K_+\right]&=2 U\,,\ \ \ \ \ \left[E_-,E_+\right]=2 U\,,\\
\left[E_+,K_-\right]&=2 V\,,\ \ \ \ \ \left[K_+,E_-\right]=2 V\,,\\
\left[U,K_\pm\right]&=\pm K_\pm\,,\ \ \ \ \left[U,E_\pm\right]=\pm E_\pm\,,\\
\left[V,K_\pm\right]&=\mp E_\pm\,,\ \ \ \ \left[U,E_\pm\right]=\mp K_\pm\,.\\
\end{aligned}\ea
However, if we directly apply this Lie-algebra to our case, the calculation is very complicated. Then in order to facilitate our calculations, we decompose this Lie algebra into a simple form
\ba\begin{aligned}\label{LJ}
L_\pm&=\frac{1}{2}\lf(K_\pm+E_\pm\rt)\,,\ \ \ \ L_0=\frac{1}{2}\lf(U-V\rt)\,,\\
J_\pm&=\frac{1}{2}\lf(K_\pm-E_\pm\rt)\,,\ \ \ \ J_0=\frac{1}{2}\lf(U+V\rt)\,.\\
\end{aligned}\ea
The non-vanish Lie-bra brackets can be written as
\ba\begin{aligned}
\left[L_-,L_+\right]&=2 L_0\,,\ \ \ \ \left[L_0, L_{\pm}\right]=\pm L_\pm\,,\\
\left[J_-,J_+\right]&=2 J_0\,,\ \ \ \ \left[J_0, J_{\pm}\right]=\pm J_\pm\,.\\
\end{aligned}\ea
These relations imply that the Lie-algebra $\math{G}$ can be decomposed into a direct product among two $su(1,1)$-algebra,$i.e.$, $\math{G}=su(1,1)\times su(1,1)$. Thus, any unitary transformation operator $U$ in the circuit $\bra{\text{TFD}(t)}=U\bra{\bar{0}}$ can be constructed by
\ba
\hat{U}=\exp\lf(\a^I L_I\rt)\exp\lf(\beta^{I} J_{I}\rt)\,.
\ea
with $I\in\{\pm,0\}$. To simplify the calculation, we redefine the corresponding  annihilation and creation operators by
\ba\begin{aligned}
A_L&=\frac{1}{\sqrt{2}}\lf(\bar{a}_L-\bar{b}_R\rt)\,,\ \ \ \ A_R=\frac{1}{\sqrt{2}}\lf(\bar{a}_R+\bar{b}_L\rt)\,,\\
B_L&=\frac{1}{\sqrt{2}}\lf(\bar{a}_L+\bar{b}_R\rt)\,,\ \ \ \ B_R=\frac{1}{\sqrt{2}}\lf(\bar{a}_R-\bar{b}_L\rt)\,.\\
\end{aligned}\ea
Then, the generators \eq{LJ} can be reexpress as
\ba\begin{aligned}
L_+=A_L^{\dag}A_R^{\dag}\,,\ \ \ L_-=A_L A_R\,,\ \ \ L_0=\frac{1}{2}\lf(A_L^\dag A_L+A_R^\dag A_R-1\rt)\,,\\
J_+=B_L^{\dag}B_R^{\dag}\,,\ \ \ J_-=B_L B_R\,,\ \ \ J_0=\frac{1}{2}\lf(B_L^\dag B_L+B_R^\dag B_R-1\rt)\,.\\
\end{aligned}\ea
Since the two parts commute, we have
\ba
\bra{T}=\hat{U}\bra{\bar{0}}=\exp\lf(\a^I L_I\rt)\bra{0}_A\otimes\exp\lf(\beta^{I} J_{I}\rt)\bra{0}_B\,.
\ea
Applying the similar procedure, we re-express the target state $\bra{T}$ into
\ba
\bra{T}=e^{-(\g^0+{\g^0}')/2}e^{\g^+ L_+}\bra{0}_A\otimes e^{{\g^+}' J_+}\bra{0}_B\,.
\ea
Then, the Fubini study line element can be obtained by
\ba
ds_{\text{FS}}^2=\frac{L}{\p}\int_0^\L dk\lf(\frac{|\pd_\s \g^+_k(\s)|^2}{(1+| \g^+_k(\s)|^2)^2}+\frac{|\pd_\s {\g^+_k}'(\s)|^2}{(1+| {\g^+_k}'(\s)|^2)^2}\rt).
\ea
where we have defined
\ba
\g^+_k(\s)=\tan\lf(\frac{\q_k(\s)}{2}\rt) e^{i \f_k(\s)}\,,\\
{\g^+_k}'(\s)=\tan\lf(\frac{\q_k'(\s)}{2}\rt) e^{i \f_k'(\s)}\,.
\ea
Then, the FS line element can be written as
\ba
ds_{\text{FS}}^2=\frac{L}{4\p}\int_0^\L dk\lf(d\q_k^2+\sin^2\q_k d\f_k^2+d\q_k'^2+\sin^2\q'_k d\f_k'^2\rt)\,.
\ea
Next, we consider the circuit from $\bra{\bar{0}}$ to the TFD state,
\ba\label{TFDstate}\begin{aligned}
\bra{TFD(t)}&=e^{\a\lf(E_++E_-\rt)} e^{z K_++z^*K_-}\bra{\bar{0}}\\
&=e^{(z+\a)L_++(z^*+\a)L_-}\bra{0}_A\otimes e^{(z-\a)J_++(z^*-\a)J_-}\bra{0}_B\,,
\end{aligned}\ea
in which
\ba
\a_k=\frac{1}{2}\arctan\lf(\frac{k}{m}\rt)\,,\ \ \ \ z_k= \q_0^{(k)} e^{-i\w_k t}\,.
\ea
Then, one can find
\ba\begin{aligned}
\q_k(\s_f)&=2\sqrt{\lf(\q_0^{(k)}\rt)^2+\a_k^2+2\q_0^{(k)} \a_k \cos \w_k t}\,,\ \ \f(\s_f)=\arctan\lf(\frac{\q_0^{(k)} \sin\w_k t}{\q_0^{(k)} \cos\w_k t+\a_k}\rt)\,,\\
\q'_k(\s_f)&=2\sqrt{\lf(\q_0^{(k)}\rt)^2+\a_k^2-2\q_0^{(k)} \a_k \cos \w_k t}\,,\ \ \f'(\s_f)=\arctan\lf(\frac{\q_0^{(k)} \sin\w_k t}{\q_0^{(k)} \cos\w_k t-\a_k}\rt)\,.
\end{aligned}\ea
Then, the circuit complexity can be obtained as
\ba\label{CFS2}
\math{C}^2_\text{FS}\lf(\bra{\bar{0}}\to\bra{TFD(t)}\rt)
=\frac{2L}{\p}\int_0^{\inf} dk\lf[\lf(\q_0^{(k)}\rt)^2+\a^2_k\rt]\,.
\ea

Note that in this case, the complexity is also time-independent. More explicitly, as denoted in \eq{TFDstate}, the complexity is time-dependent for one part of the system, $e.g.$, $A$ part, while the time variance cancels out when considering the whole system. However, unlike the circuit start from the zero temperature reference state, this circuit complexity start from the NSE state is UV divergent. From \eq{CFS2}, we can see that the complexity for the TFD state also shares the same divergence structure with the Dirac vacuum state. Moreover, in order to obtain a UV structure as the holographic complexity, we define an $L^2$ norm complexity as $\math{C}^{(2)}_\text{FS}=\math{C}^2_\text{FS}$. With similar consideration as Sec.\ref{CCTFD}, we also define the relative complexity as
\ba\begin{aligned}\label{rc2}
\D\math{C}^{(2)}_\text{FS}\lf(\bra{TFD(t)},\bra{0}\rt)&=\math{C}^2_\text{FS}\lf(\bra{\bar{0}}\to\bra{TFD(t)}\rt)
-\math{C}^2_\text{FS}\lf(\bra{\bar{0}}\to\bra{0}\rt)\\
&=\frac{2L}{\p}\int_0^{\inf} dk\lf(\q_0^{(k)}\rt)^2\,,
\end{aligned}\ea
which gives the same expression as \eq{Ck} for $n=2$, although it is necessary to recast the expression \eq{Ck} as
\ba\begin{aligned}
\math{C}_\k\lf(\bra{\Omega_0}\to \bra{\text{TFD}(t)}\rt)&=\frac{c_fV}{(2\p)^{n-1}}\int d^{n-1}k |\theta_0^{(k)}|^\k \\
&=\math{C}_\k\lf(\bra{\Omega_0}\to \bra{\text{TFD}(t)}\rt)-\math{C}_\k\lf(\bra{\Omega_0}\to \bra{\Omega_0}\rt)\\
&=\Delta\math{C}_\k\lf( \bra{\text{TFD}(t)},\bra{\Omega_0}\rt).
\end{aligned}\ea

To conclude this section, we have investigated the complexity of two circuits by using the Fubini-Study metric approach, one is from the zero temperature state to the TFD state, the other is from NSE state to the TFD state. We found that both expression \eq{COFS} and \eq{rc2} share the similar formula with \eq{Ck}, especially, if we define $L^2$ norm, $i.e.,{}\math{C}^{(2)}_\text{FS}=\math{C}^2_\text{FS}$, in the Fubini-Study approach,  the results are exactly the same with the continuum case of the complexity of $2N$ fermionic oscillators for the measure $\kappa=2$ evaluated by the Nielsen's approach. It might be interesting to further investigate the property of $L^2$ norm for both methods in the interacting fermionic field theory.

\section{Conclusion}\label{conclusion}
In this paper, motivated by recent development of circuit complexity for TFD states in the free scalar theory, we have generalized some of those results to the time-dependent TFD states in free, non-interacting fermionic system. Our starting point is to compare the circuit complexity for a time-dependent TFD states in a free fermionic theory with the results of the complexity of the boundary TFD states in the context of holography. Recent developments have provided us with two computable approaches to evaluate the circuit complexity in the context of free field theory--Nielsen's geometry approach and Fubini-study metric approach. We also compared the results of those two methods and found out that the circuit complexity of those two methods shares some extraordinary similarities.

Following the ideas of Nielsen and collaborators, the circuit which prepares the desired state lies in a Riemannian geometry of the underlying Lie group, in our case SO(4$N,\mathbb{R}$) group, with appropriate boundary condition. The optimal circuits become geodesics in this geometry. It is noticeable that we are considering a whole class of circuits which gives us the desired fermionic TFD state. To avoid unnecessary computation, we express the Gaussian states into covariance matrix form for evaluating the circuit complexity. We began with the 2$N$ decoupled harmonic oscillators for the whole fermionic TFD state and then extended the analysis to the continuum case to study the complexity of formation. We found out that $\Delta\math{C}_{\kappa}$ is positive and also UV finite for both $L^1$ norm and $L^2$ norm, which is very similar to the holographic result. Furthermore, If we consider a massless case, $\Delta\math{C}_{\kappa=1}$ is proportional to the entanglement entropy between the left and right CFTs as found with holographic complexity when the dimension is greater than 3, however, $\Delta\math{C}_{\kappa=2}$ is inversely proportional to the entanglement entropy in contrast to the holographic counterpart. For two dimensional free Dirac field, we first considered the massless case, our results \eq{0C2}\eq{0C1} show that the complexities of formation for both measures increase with decrease with the growth of temperature.  Then, for the massive case, the complexities of formation for both measures start from zero, increase with $\tilde{T}$, and after reach their positive maximum values their figures decrease to $-\inf$. Results from both cases are different with the results from holography. As for the time-dependence of the TFD state, the complexities for both measures in the massless case increase with time at a short period then saturate at a constant. In the massive case, the complexities show damped oscillations which leads to a quick saturation. Again, those features are very different from the late-time linear growth found for holographic complexity. Since the late-time growth of holographic complexity is a crucial property of the CA and CV conjectures, we might be disappointed that our analysis did not recover this feature. However, one must aware that the fermionic QFT for which we are analyzing the complexity is a free theory, while the boundary CFTs which are described by holographic complexity are strongly coupled system. Furthermore, the states we are analyzed is confined to the space of Gaussian states which is just the subspace of the full Hilbert space. Our work here just provides some basics calling for the future to investigate more complicated cases, such interacting fermionic field or the full Hilbert space.

 In the last section, we investigated the circuit complexity generated by the Fubini-Study metric approach. First we studied the complexity from the zero temperature reference state to the fermionic TFD state and found out that the square of this result $\math{C}_{FS}(\bra{\Omega_0}\to\bra{\text{TFD}(t)})$ \eq{COFS} shares the same expression with \eq{Ck} when $\kappa=2$ by using Nielsen's approach in a continuum fermionic field theory. Then we changed our reference to the NSE reference states. Again, we discovered that the relative complexity \eq{rc2} shares the same formula with \eq{Ck}. These similarities shared with the Nielsen's approach for a specific measure might just be a coincidence, or it has more profound mathematical or physical meanings. This needs researchers to further explore in the future.

\begin{acknowledgments}
J.Jiang is partially supported by Natural Science Foundation of China (NSFC) with Grant No.11375026, and 11775022. X.J.Liu is supported in part by the NSFC with Grant No. 11475023 and No. 11875006.
\end{acknowledgments}

\appendix
\section{Transformation matrix for the time-dependent TFD state}\label{A}
In this appendix, we compute the transformation matrix of time-dependent TFD state, i.e.,
\ba\label{UabTFD}
\hat{U}(\q)^{\dag}\x^a\hat{U}(\q)=U(\q)^a{}_b\x^b\,.
\ea
with
\ba
\hat{U}(\q)=e^{\q\hat{K}(t)}\,.
\ea
We can define the generator $K^a{}_b$ of the transformation matrix $U^a{}_b(\q)$, such that
\ba
U(\q)=e^{\q K}\,.
\ea
Then, we have
\ba
[\x^a, \hat{K}]=K^a{}_b\x^b\,,
\ea
Using Eq.\eq{mmkt}, one can obtain
\ba\begin{aligned}
\lf[Q_L,\hat{K}\rt]&=Q_R \cos \w t-P_R \sin\w t\,,\\
\lf[Q_R,\hat{K}\rt]&=-Q_L \cos \w t+P_L \sin\w t\,,\\
\lf[P_L,\hat{K}\rt]&=-P_R \cos \w t-Q_R \sin\w t\,,\\
\lf[P_R,\hat{K}\rt]&=P_L \cos \w t+Q_L \sin\w t\,,
\end{aligned}\ea
which gives
\ba
K=
\left(
\begin{array}{cccccccc}
 0 & \cos \w t & 0  & -\sin\w t \\
 -\cos \w t & 0 & \sin\w t & 0  \\
 0 & -\sin\w t & 0 & -\cos \w t \\
 \sin\w t & 0 & \cos \w t & 0 \\
\end{array}
\right)\,.
\ea
This generator matrix can be diagonalized by the matrix $S$, i.e., $K=S K_0 S^{-1}$ with $K_0=\text{diag}\{i,i,-i,-i\}$ and
\ba
S=
\left(
\begin{array}{cccc}
 i \csc \omega t & -\cot  \omega t & -i \csc  \omega t  & -\cot  \omega t \\
 -\cot \omega t & -i \csc  \omega t & -\cot  \omega t  & i \csc  \omega t \\
 0 & 1 & 0 & 1 \\
 1 & 0 & 1 & 0 \\
\end{array}
\right)\,.
\ea
The transformation matrix can be obtained by
\ba
e^{\q K}=\sum_{\inf}^{n=0}\frac{\q^n}{n!}K^n=S\sum_{\inf}^{n=0}\frac{\q^n}{n!}K_0^nS^{-1}=S e^{\q K_0} S^{-1}\,,
\ea
in which $e^{\q K_0}$ can be obtained as $e^{\q K_0}=\text{diag}\{e^{i \q},e^{i \q},e^{-i \q},e^{-i \q}\}$. Then, we can obtain
\ba\label{Uq}
U(\q)=
\left(
\begin{array}{cccc}
 \cos \theta & \cos \omega t\sin \theta  & 0 & -\sin \theta  \sin  \omega t \\
 -\cos  \omega t  \sin \theta  & \cos \theta  & \sin \theta  \sin  \omega t & 0 \\
 0 & -\sin \theta \sin \omega t & \cos \theta & -\cos  \omega t \sin \theta  \\
 \sin \theta  \sin \omega t & 0 & \cos  \omega t \sin \theta  & \cos \theta  \\
\end{array}
\right)\,.
\ea
\section{FS metric derivation for the Fermionic TFD state}\label{B}
In this appendix, we present the derivation of \eq{CFS} for the TFD state of a simple Fermionic harmonic oscillator. The generator set form a Lie-algebra $su(1,1)$. Then, By using the decomposition formula
\ba
e^{\a^I K_I}=e^{\g^+K_+}e^{\g^0K_0}e^{\g^-K_-}
\ea
with
\ba\begin{aligned}
\g^{\pm}&=\frac{2\a^\pm \sinh \X}{2\X\cosh\X-\a^0 \sinh\X}\,,\\
\g^0&=\lf(\cosh \X-\frac{\a^0}{2\X}\sinh\X\rt)^{-2}\,,\\
\X^2&=\frac{\lf(\a^0\rt)^2}{4}-\a^+\a^-\,,
\end{aligned}\ea
which is taken from appendix 11.3.3 of Ref.\cite{Chapman:2017}. Then, by the normalization, the target state can be written as
\ba\begin{aligned}
\bra{\y}&=\lf(1+|\g^+|^2\rt)^{-1/2}e^{\g^+K_+}\bra{0}_L\otimes\bra{0}_R\\
&=\lf(1+|\g^+|^2\rt)^{-1/2}e^{\g^+a_L^\dag a_R^\dag}e^{\g^+b_L^\dag b_R^\dag}\bra{0}_L\otimes\bra{0}_R\\
&=\lf(1+|\g^+|^2\rt)^{-1/2}\lf[\bra{0,0}_L\oplus\bra{0,0}_R-\lf(\g^+\rt)^2\bra{1,1}_L\oplus\bra{1,1}_R\right.\\
&\left.+\g^+\lf(\bra{0,1}_L\oplus\bra{0,1}_R+\bra{1,0}_L\oplus\bra{1,0}_R\rt)\rt]\,,
\end{aligned}\ea
where we denote
\ba
a^\dag_\L\bra{0}_\L=\bra{1,0}_\L\,,\ \ \ \ b^\dag_\L\bra{0}_\L=\bra{0,1}_\L\,.
\ea
Then, the FS line element
\ba
ds^2_\text{FS}=\langle\d\y\bra{\d\y}-\langle\y\bra{\d\y}\langle\d\y\bra{\y}
\ea
can be directly obtained
\ba
ds^2_\text{FS}=\frac{|\d\g^+|^2}{(1+|\g^+|^2)^2}\,.
\ea

\end{document}